\documentclass[aps,prd,preprintnumbers,groupedaddress,nofootinbib,amssymb,notitlepage,eqsecnum]{revtex4-2}
\usepackage{here}
\usepackage[dvipdfmx]{graphicx}
\usepackage{amsmath,amsthm,amssymb}
\usepackage{bm}
\usepackage{color}

\usepackage{amsfonts}
\usepackage{dcolumn}
\usepackage{hyperref}
\allowdisplaybreaks[1]
\usepackage{stackengine}

\newcommand{\be}{\begin{equation}}  
\newcommand{\ee}{\end{equation}}
\newcommand{\ba}{\begin{eqnarray}}
\newcommand{\ea}{\end{eqnarray}}

\newcommand{\rd}{{\rm d}}

\newcommand{\bem}{\begin{bmatrix}}
\newcommand{\eem}{\end{bmatrix}}
\newcommand{\Mpl}{M_{\rm Pl}}

\newcommand{\mL}{\mathcal{L}}

\allowdisplaybreaks

\begin{document}

\preprint{YITP-23-63, WUCG-23-06}

\title{Excluding static and spherically symmetric black holes 
in Einsteinian cubic gravity \\
with unsuppressed higher-order curvature terms}

\author{Antonio De Felice$^{1}$ and Shinji Tsujikawa$^{2}$}

\affiliation{
$^1$Center for Gravitational Physics and Quantum Information, 
Yukawa Institute for Theoretical Physics, Kyoto University, 
606-8502, Kyoto, Japan\\
$^2$Department of Physics, Waseda University, 3-4-1 Okubo, Shinjuku, Tokyo 169-8555, Japan}

\begin{abstract}

Einsteinian cubic gravity is a higher-order gravitational 
theory in which the linearized field equations of motion match Einstein's equations on a maximally symmetric background. 
This theory allows the existence of a static and spherically symmetric 
black hole solution where the temporal and radial metric components are equivalent to each other ($f=h$), with a modified Schwarzschild 
geometry induced by cubic curvature terms. 
We study the linear stability of the static and spherically symmetric vacuum 
solutions against odd-parity perturbations without dealing with Einsteinian cubic gravity as an effective field theory where the cubic curvature 
terms are always suppressed relative to the Ricci scalar.
Unlike General Relativity containing one dynamical perturbation, 
Einsteinian cubic gravity has three propagating degrees of freedom in the odd-parity sector. We show that at least one of those dynamical perturbations always behaves as a ghost mode. 
We also find that one dynamical degree of freedom has a negative sound speed squared $-1/2$ for the propagation of high angular momentum modes.
Thus, the static and spherically symmetric hairy black hole solutions 
realized by unsuppressed cubic curvature terms relative to 
the Ricci scalar are excluded by ghost and 
Laplacian instabilities.

\end{abstract}

\date{\today}


\maketitle

\section{Introduction}
\label{introsec}

General Relativity (GR) has been successful in describing the gravitational interaction between submillimeter and solar-system scales \cite{Hoyle:2000cv,Adelberger:2003zx,DeFelice:2010aj,Clifton:2011jh,Will:2014kxa}. 
At extremely small distances close to the Planck length, however, it is expected that GR is replaced by a quantum theory of gravity with an ultraviolet completion. The attempt for the construction of a power-counting renormalizable gravitational theory was advocated by Stelle \cite{Stelle:1976gc} by taking into account quadratic-order curvatures to the Einstein-Hilbert action. In string theory, the low-energy effective action also contains quadratic curvature corrections known as a Gauss-Bonnet (GB) term \cite{Zwiebach:1985uq,Gross:1986mw}. 
Moreover, higher-order curvature terms have played a prominent role in conformal field theory with holography \cite{Hofman:2009ug,Myers:2010ru,Myers:2010jv,Oliva:2010eb}.

In gravitational theories where the field equations of motion contain derivatives higher than second order in the metric tensor $g_{\mu \nu}$, the system can be unstable due to the emergence of Ostrogradski instability \cite{Ostrogradsky:1850fid,Woodard:2015zca} associated with a Hamiltonian unbounded from below.
To avoid such a problem, Lanczos \cite{Lanczos:1938sf} and Lovelock \cite{Lovelock:1971yv} constructed gravitational theories with second-order field equations of motion for general, curved backgrounds. Exploiting polynomial functions of the Riemann curvature tensors in four-dimensional spacetime endowed with four-dimensional diffeomorphism invariance, Lovelock showed that the symmetric and divergence-free tensors $A_{\mu \nu}$, which depend on $g_{\mu \nu}$ and its derivatives up to second order, are expressed by a linear combination of the Einstein tensor $G_{\mu \nu}$ and the metric tensor $g_{\mu \nu}$. 
In spacetime dimensions higher than four, there is the quadratic-order GB curvature scalar affecting the spacetime dynamics. 
In four dimensions, the GB term corresponds to an Euler density which does not contribute to the field equations of motion. 
To extract its effect in four dimensions, the GB term needs to be coupled to other degrees of freedom (DOFs) such as scalar or vector fields \cite{Metsaev:1987zx,Antoniadis:1993jc,Gasperini:1996fu,Kanti:1995vq,
Torii:1996yi,Kawai:1998ab,Cartier:1999vk,Tsujikawa:2002qc,Aoki:2023jvt}.

If we consider cubic-order curvature combinations constructed from the Riemann tensors, there is also an additional Euler density in the action corresponding to the surface integral in four dimensions \cite{Lovelock:1971yv}.
According to the Lovelock theorem, there are no other nontrivial cubic-order terms keeping the field equations of motion up to second order in general, curved backgrounds. If we consider some specific spacetime, however, it is possible to construct nontrivial cubic-order theory whose graviton spectrum shares a similar property to that in GR. On a maximally symmetric background, which includes the Minkowski and de Sitter spacetimes, there is a unique cubic combination ${\cal P}$ which keeps the structure of linearized perturbation equations of motion in Einstein gravity \cite{Bueno:2016xff}. 
In this theory, which is dubbed {\it Einsteinian cubic gravity} (ECG), the cubic curvature term can give rise to derivatives higher than second order in general, curved backgrounds, so the spacetime dynamics is generally different from that in GR \cite{Bueno:2016ypa}.

On a static and spherically symmetric (SSS) background in a vacuum configuration, ECG admits the existence of a black hole (BH) solution whose temporal and radial metric components (denoted as $f$ and $h$, respectively) coincide with each other \cite{Hennigar:2016gkm,Bueno:2016lrh,Bueno:2017sui,Hennigar:2018hza} 
(see also Refs.~\cite{Adair:2020vso,Frassino:2020zuv,Burger:2019wkq,Cano:2019ozf}).
Since $f$ and $h$ are affected by the cubic curvature term, the background geometry is still different from the Schwarzschild BH solution. One can also construct a cubic gravity theory by imposing the condition $f=h$ (up to a time reparametrization freedom in $f$) on the SSS background \cite{Hennigar:2017ego,Bueno:2019ltp,Bueno:2019ycr}. 
This construction allows the existence of the Lagrangian ${\cal P}$ mentioned above as well as two additional cubic combinations ${\cal C}$ and ${\cal C}'$ defined in Ref.~\cite{Hennigar:2017ego}. 
As we will see later in Sec.~\ref{scasec}, both ${\cal C}$ and ${\cal C}'$ vanish for $f=h$ and hence only the Lagrangian ${\cal P}$ contributes to the single differential equation for $f~(=h)$.

If we apply ECG to the cosmological dynamics on the Friedmann-Lema\^{i}tre-Robertson-Walker (FLRW) spacetime, the field equations of motion arising from ${\cal P}$ contain derivatives higher than second order. On the FLRW background, there is a unique combination ${\cal P}-8{\cal C}$ that leads to the second-order Friedmann equation \cite{Arciniega:2018tnn,Arciniega:2018fxj}. 
This theory--dubbed {\it cosmological Einsteinian cubic gravity} (CECG)-- was applied to the dynamics of inflation and late-time cosmological epochs \cite{Arciniega:2018tnn,Arciniega:2018fxj,Cisterna:2018tgx,Erices:2019mkd,Quiros:2020uhr,BeltranJimenez:2020lee,Cano:2020oaa}.
On an exact de Sitter background, tensor perturbations with two polarized modes propagate as in GR without pathological behavior. If one considers a spatially homogeneous Bianchi type I manifold close to the isotropic de Sitter spacetime, however, there are three dynamical propagating DOFs associated with linear perturbations in the odd-parity sector.
In Ref.~\cite{Pookkillath:2020iqq}, it was shown that, in the regime of small anisotropies, such theory possesses at least one ghost mode as well as short-time-scale tachyonic instability. Hence CECG cannot be used to describe a viable geometric inflationary scenario. 

Given that there is an instability problem of CECG on the anisotropic cosmological background, we are now interested in the linear stability of SSS vacuum solutions in ECG. For this purpose, we do not deal with ECG as a trivial effective field theory (EFT) where the cubic Lagrangian is always strongly suppressed relative to the Einstein-Hilbert term.
We consider odd-parity perturbations according to the Regge-Wheeler formulation \cite{Regge:1957td} without necessarily imposing the condition $f=h$. 
We show that the Lagrangian ${\cal P}$ in ECG gives rise to three propagating DOFs in the odd-parity sector. 
We find that there is at least one ghost mode and that the squared propagation speed of one of the dynamical perturbations is negative for large multipoles in the regime where the effective mass of perturbations is below 
the Planck mass, or the cutoff of the theory $M$.
The presence of these ghost and Laplacian instabilities excludes the SSS vacuum solutions in ECG with unsuppressed higher-order curvature terms, including the BH solution with $f=h$. 

This paper is organized as follows. 
In Sec.~\ref{scasec}, we revisit the SSS BH solution present in ECG.
In Sec.~\ref{persec}, we derive the second-order action of odd-parity perturbations and show how the ghost and Laplacian instabilities arise for the large frequency and momentum modes.
Sec.~\ref{consec} is devoted to conclusions.

\section{Cubic gravity}
\label{scasec}

General cubic gravity theories consist of a combination of cubic products of the Riemann tensor $R_{\alpha \beta \gamma \delta}$, Ricci tensor $R_{\mu \nu}$, and Ricci scalar $R$.  
We consider the following eight cubic Lagrangians: 
\ba
& &
\mL_1={{R_{\alpha}}^{\beta}}{}_{\gamma}{}^{\delta}
{{R_{\beta}}^{\mu}}{}_{\delta}{}^{\nu}
{{R_{\mu}}^{\alpha}}{}_{\nu}{}^{\gamma}\,,\qquad
\mL_2=R_{\alpha \beta}{}^{\gamma \delta}
R_{\gamma \delta}{}^{\mu \nu}
R_{\mu \nu}{}^{\alpha \beta}\,,\qquad
\mL_3=R_{\alpha \beta \gamma \delta}
R^{\alpha \beta \gamma}{}_{\mu}
R^{\delta \mu}\,,\qquad 
\mL_4=R R_{\alpha \beta \gamma \delta} 
R^{\alpha \beta \gamma \delta}\,, \nonumber \\
& &
\mL_5=R_{\alpha \beta \gamma \delta} 
R^{\alpha \gamma} R^{\beta \delta}\,,\qquad
\mL_6=R_{\alpha}{}^{\beta} R_{\beta}{}^{\gamma} 
R_{\gamma}{}^{\alpha}\,,\qquad 
\mL_7=R R_{\alpha \beta} R^{\alpha \beta}\,,\qquad 
\mL_8=R^3\,.
\ea
Taking into account the Einstein-Hilbert term $\Mpl^2 R/2$, where $\Mpl$ is the reduced Planck mass, we express the total action as 
\be
{\cal S}=\frac{M_{\rm pl}^2}{2} \int {\rm d}^4 x \sqrt{-g} 
\left( R+\sum_{i=1}^{8} c_i \mL_i \right)\,,
\label{Saction}
\ee
where $g$ is a determinant of the metric tensor $g_{\mu \nu}$, and $c_i$'s are constants.

ECG is constructed to possess a transverse and massless graviton spectrum as in GR on a maximally symmetric background \cite{Bueno:2016xff}. 
Requiring also that the relative coefficients of different curvature terms are the same in all dimensions, there is a unique combination ${\cal P}$ which is neither trivial nor topological in four dimensions. 
This nontrivial cubic interaction corresponds to the choice of coefficients $c_1=12$, $c_2=1$, $c_5=-12$, and $c_6=8$, i.e., 
\be
{\cal P}= 
12\mL_1+\mL_2-12\mL_5+8\mL_6\,.
\label{P}
\ee
In general curved spacetime including the SSS background as well as the FLRW background, the equations of motion following from the Lagrangian ${\cal P}$ 
are higher than the second order. In such cases, higher-order derivatives can induce extra DOFs in comparison to those in GR.

In Ref.~\cite{Hennigar:2017ego}, the authors took a different approach to the construction of cubic gravity theories (dubbed generalized quasi-topological gravity) by demanding that the vacuum SSS solution is fully characterized by a single field equation. In general, the line element on the SSS background is given by 
\be 
\rd s^2=-f(r) \rd t^{2} +h^{-1}(r) 
\rd r^{2}+ r^{2} \left( \rd \theta^{2}
+\sin^{2}\theta\,\rd\varphi^{2} 
\right)\,,
\label{background}
\ee
where $f$ and $h$ are functions of the radial coordinate $r$. 
If $f(r)$ is proportional to $h(r)$ such that $f(r)=N h(r)$, 
where $N$ is a constant, the time and radial components of 
the field equations coincide with each other. 
This gives the following constraints 
among the coefficients in Eq.~(\ref{Saction}): 
\be
c_4=\frac{3c_1-36c_2-14c_3}{56},\quad 
c_5=-\frac{3c_1+48c_2+14c_3}{7},\quad 
c_7=\frac{6c_1+96c_2+14c_3-21c_6}{28},\quad
c_8=-\frac{3c_1+20c_2-7c_6}{56}.
\ee
The six-dimensional Euler density ${\cal X}_6$ corresponds to the coefficients $c_1=8$, $c_2=-4$, $c_3=24$, and $c_6=-16$, but this is a topological term that does not affect the field equations of motion. 
One of the remaining three cubic interactions is the Lagrangian ${\cal P}$ given by Eq.~(\ref{P}). 
There are also two additional terms
\ba
{\cal C} &=& \mL_3-\frac{1}{4}\mL_4-2\mL_5
+\frac{1}{2}\mL_7\,,\label{C}\\
{\cal C}' &=& \mL_6-\frac{3}{4}\mL_7+\frac{1}{8}\mL_8\,,
\label{Cd}
\ea
which correspond to the coefficients $c_1=0$, $c_2=0$, $c_3=1$, $c_6=0$ and
$c_1=0$, $c_2=0$, $c_3=0$, $c_6=1$, respectively. 
As we will see below, for $f=h$, the contributions of the Lagrangians ${\cal C}$ and ${\cal C}'$ to the field equations of motion vanish. 
On the FLRW spacetime, the Lagrangian ${\cal P}$ gives rise to the Friedmann equation higher than second-order. 
The specific combination ${\cal P}-8{\cal C}$ leads to the second-order field equations on the FLRW background \cite{Arciniega:2018tnn,Arciniega:2018fxj}. 
If we apply this latter cubic theory to inflation, the presence of small anisotropies close to the de Sitter background generates the instability of cosmological solutions \cite{Pookkillath:2020iqq}. 

In both ECG and CECG, the dynamical DOFs are more than those 
in GR (two tensor polarizations) around general, curved backgrounds.  
The property that the propagating DOFs around the maximally 
symmetric background degenerate to those in GR implies that 
the disappearing DOFs are, in general, strongly coupled \cite{BeltranJimenez:2020lee,Jimenez:2023esa}. 
In such theories, the maximally symmetric background corresponds 
to a singular surface at which the coefficients of higher-order 
kinetic terms appearing in the perturbation equations of motion are degenerate.
This degeneracy leads to the divergence of couplings of the canonically 
normalized fields, which is a signal of the strong coupling. 
On the background different from the maximally symmetric spacetime, 
some pathological behavior like the instability of perturbations usually
arises as it happens for the cosmological 
background \cite{Pookkillath:2020iqq,BeltranJimenez:2020lee}.  
In this paper, we would like to study whether this is also the case 
for BHs on the SSS background.

On the SSS background given by the line element (\ref{background}), lets us consider cubic gravity theories given by the action 
\be
{\cal S}=\frac{M_{\rm pl}^2}{2} \int {\rm d}^4 x \sqrt{-g} 
\left( R+\alpha {\cal P}+\kappa {\cal C}
+\mu {\cal C}' \right)\,,
\label{action}
\ee
where $\alpha$, $\kappa$, $\mu$ are constants. 
Then, the quantities (\ref{P}), (\ref{C}), and (\ref{Cd}) reduce, respectively, to 
\ba
{\cal P} &=& \frac{1}{f^4 r^4} [3f'^4 h^3r^2 
- 3f^3 h' (2f'' hr + f' h' r - 2f' h)(h'r - 2h + 2)
- 6 f'^2 f h^2 r \{ f'' h r + f'( h'r + h - 1)\} \nonumber \\
& &+ 6r f' f^2 h \{ f'h'( h'r + h) + f'' h( h'r + 2h - 2) \}]\,,\\
{\cal C} &=& -\frac{3}{8 f^4 r^4} \left( f' h-h' f \right)^2 
\left[ (2f''fh-hf'^2+f'h' f)r^2-4f^2 (h-1) \right]\,,\\
{\cal C}' &=& \frac{1}{2} {\cal C}\,,
\ea
where a prime represents the derivative with respect to $r$.
If $f$ is equal to $h$, both ${\cal C}$ and ${\cal C}'$ vanish. Varying the action (\ref{action}) with respect to $f$ and $h$, we obtain the third-order differential equations for $h$ and $f$, respectively\footnote{If we relax the assumption of staticity, it can be shown that for this theory the Birkhoff theorem does not hold in general. Therefore, there could be other solutions that in principle have some relevance but will be in general time-dependent. We will not investigate their existence, however, as we will already set strong constraints on the theory just by looking at their static limit.}. 
Setting $f(r)=h(r)$, the two differential equations coincide with each other, giving
\be
r f'+f-1+\frac{6\alpha}{r^3} [ rf'^2 (4f-1)
+rf \{ r^2 f''^2+4f'' (f-1)-2r f'''(f-1) \}
+f' f(r^3 f'''-4r^2 f''-4f+4)]=0\,.
\label{difeq}
\ee
For $\alpha=0$, the solution to Eq.~(\ref{difeq}) is $f=h=1-r_h/r$, where $r_h$ is an integration constant corresponding to the horizon radius.
For $\alpha \neq 0$, there are corrections to the Schwarzschild metric. 
At large distances away from the horizon ($r \gg r_h$), we derive the solution to Eq.~(\ref{difeq}) under the following expansion
\be
f(r)=h(r)=1-\frac{r_h}{r}
+\sum_{i=2}c_i \left( \frac{r_h}{r} \right)^{i}\,,
\label{fh}
\ee
where $c_i$'s are constants.
Substituting Eq.~(\ref{fh}) into Eq.~(\ref{difeq}) and computing the coefficients at each order, we obtain $c_2=c_3=c_4=c_5=0$, $c_6=54\alpha/r_h^4$,
and $c_7=-46 \alpha/r_h^4$.
Then, the leading-order correction to $f$ arises at sixth order in the expansion (\ref{difeq}), so that\footnote{Here we do not study whether the series converges, as in any case, the solution loses its validity for large enough values of $r$, but we assume that it can be considered as a good approximation in the physical range of $r$ of interest to the real numerical solution having suitable asymptotically flat boundary conditions.} 
\be
f(r)=h(r)=1-\frac{r_h}{r}
+\frac{54\alpha r_h^2}{r^6}\left[1
+{\cal O}\left( \frac{r_h}{r} \right) \right]\,.
\label{fh2}
\ee

The solution (\ref{fh2}) has been derived by requiring that $f(r)=h(r)$ without explicitly imposing any requirement on the value of $\alpha$. 
The same solution also follows without assuming the condition $f(r)=h(r)$ in ECG with 
\be
{\cal P} \neq 0\,,\qquad {\cal C}=0\,,\qquad 
{\cal C}'=0\,.
\ee
In this case, we write the large-distance solutions as
\be
f(r)=1-\frac{r_h}{r}
+\sum_{i=2}c_i \left( \frac{r_h}{r} \right)^{i}\,,\qquad 
h(r)=1-\frac{r_h}{r}
+\sum_{i=2}d_i \left( \frac{r_h}{r} \right)^{i}\,.
\label{fhex}
\ee
We substitute Eq.~(\ref{fhex}) into the third-order differential equations of $f$ and $h$ and obtain the coefficients $c_i$ and $d_i$.
This gives rise to the same coefficients $c_i$ derived above, with $c_i=d_i$, so that the large-distance solution is again given by Eq.~(\ref{fh2}). 
Hence the BH solution with $f(r)=h(r)$ generically arises in ECG.

\section{Odd-parity perturbations on the SSS background}
\label{persec}

We study the stability of SSS vacuum solutions in ECG given by the action 
\be
{\cal S}=\frac{M_{\rm pl}^2}{2} \int {\rm d}^4 x 
\sqrt{-g} \left( R+\alpha {\cal P} \right)\,.
\label{action2}
\ee
In our analysis, we do not restrict ourselves to the EFT regime where $\alpha {\cal P}$ is always suppressed relative to $R$. 
The modification to the Schwarzschild BH in GR can be significant by allowing for the possibility that $\alpha {\cal P}$ can be as large as $R$. 
This is analogous to Starobinsky's model given by the Lagrangian $L=R+\beta R^2$ \cite{Starobinsky:1980te}, where the cosmic acceleration occurs in the regime $\beta R^2 \gtrsim R$. If we stick to the EFT regime with $\beta R^2 \ll R$, 
one cannot accommodate the physics of inflation driven by 
the quadratic curvature term.

On the background (\ref{background}), we decompose the metric tensor into $g_{\mu \nu}=g_{\mu \nu}^{(0)}+h_{\mu \nu}$, where $g_{\mu \nu}^{(0)}$ and $h_{\mu \nu}$ correspond to the background and perturbed parts, respectively.
Although we are primarily interested in the stability of the BH solution (\ref{fh2}), we do not impose the condition $f=h$ for the background metric from the beginning.
Under the rotation in the $(\theta, \varphi)$ plane, the metric perturbations $h_{\mu \nu}$ can be separated into odd- and even-parity modes \cite{Regge:1957td,Zerilli:1970se}.
Expanding $h_{\mu \nu}$ in terms of the spherical harmonics $Y_{l m} (\theta, \varphi)$, the odd- and even-modes have parities $(-1)^{l+1}$ and $(-1)^l$, respectively. 
In the odd-parity sector, the components of $h_{\mu \nu}$ are given by \cite{DeFelice:2011ka,Motohashi:2011pw,Kobayashi:2012kh,Kase:2014baa,Kase:2021mix}
\ba
& &
h_{tt}=h_{tr}=h_{rr}=0\,,\qquad 
h_{ab}=0\,,
\nonumber \\
& &
h_{ta}=\sum_{l,m}Q(t,r)E_{ab}
\nabla^bY_{lm}(\theta,\varphi)\,,
\qquad
h_{ra}=\sum_{l,m}W(t,r)E_{ab} 
\nabla^bY_{lm}(\theta,\varphi)\,,\nonumber \\
& &
h_{ab}=
\frac{1}{2} \sum_{l,m}
U (t,r) \left[
E_{a}{}^c \nabla_c \nabla_b Y_{lm}(\theta,\varphi)
+ E_{b}{}^c \nabla_c \nabla_a Y_{lm}(\theta,\varphi)
\right]\,,
\ea
where $Q$, $W$, and $U$ depend on $t$ and $r$, 
and the subscripts $a$ and $b$ denote either $\theta$ 
or $\varphi$. 
$E_{ab}$ is an antisymmetric tensor with nonvanishing components 
$E_{\theta \varphi}=-E_{\varphi \theta}=\sin \theta$. 
In a strict sense, we should write subscripts $l$ and $m$ 
for the variables $Q$, $W$, and $U$, but we omit 
them for brevity. 
Under the gauge transformation $x_{\mu} \to x_{\mu}+\xi_{\mu}$, 
where $\xi_t=0$, $\xi_r=0$, and $\xi_a=\sum_{l,m} \Lambda(t,r) 
E_{ab} \nabla^b Y_{lm} (\theta, \varphi)$, 
metric perturbations transform as $Q \to Q+\dot{\Lambda}$, 
$W \to W+\Lambda'-2\Lambda/r$, and $U \to U+2\Lambda$. 
There are several gauge-invariant combinations like 
\be
\hat{W} \equiv W-\frac{1}{2}U'+\frac{1}{r}U\,,\qquad
\hat{Q} \equiv Q-\frac{1}{2}\dot{U}\,.
\ee
We fix the residual gauge DOF by choosing
\be
U=0\,,
\ee
under which $\hat{W}=W$ and $\hat{Q}=Q$.

For the purpose of expanding the action (\ref{action2}) up to second order in odd-parity perturbations, we will focus on the axisymmetric modes ($m=0$) without loss of generality because nonaxisymmetric modes ($m \neq 0$) can be restored under a suitable rotation.
For the integrals with respect to $\theta$ and $\varphi$, we exploit the following properties
\ba
& &
\int_0^{2\pi}{\rm d} \varphi \int_0^{\pi} 
{\rm d} \theta\,
(Y_{l0, \theta})^2 \sin \theta=l(l+1)\,,\qquad
\int_0^{2\pi}{\rm d} \varphi \int_0^{\pi} 
{\rm d} \theta\, 
\left[ \frac{(Y_{l0, \theta})^2}{\sin \theta}+
(Y_{l0, \theta \theta})^2 \sin \theta \right]
=l^2(l+1)^2\,,\nonumber \\
& &
\int_0^{2\pi}{\rm d} \varphi \int_0^{\pi} 
{\rm d} \theta\,\left[ \left( \frac{2}{\sin \theta}
-\frac{3}{\sin^3 \theta} \right)(Y_{l0, \theta})^2
+\frac{3}{\sin \theta}(Y_{l0, \theta \theta})^2
+\sin \theta\, (Y_{l0, \theta \theta \theta})^2 
\right]=\frac{1}{4}l^2 (l+1)^2 (4l^2-2l-3)\,.
\ea
After the integration with respect to $\theta$ and $\varphi$, 
the second-order action of perturbations, which is expressed in the form 
${\cal S}^{(2)}=\int {\rm d}t {\rm d}r\,L$, contains time derivatives 
such as $\ddot{W}^2$, $\dot{Q}'^2$, and $\dot{Q}^2$. 
They can be factored out as 
\be
L_K \equiv \frac{3\alpha M_{\rm pl}^2 
\sqrt{h}[2f'hr +f(2-2h-rh')]l (l+1)}
{2f^{5/2}r^2}
\left( \ddot W -{\dot Q}' +\frac{2\dot{Q}}{r} 
\right)^2\,,
\label{LK}
\ee
where $L_K \in L$. 
Then, the rest of the Lagrangian $L-L_K$ does not contain products like $\ddot{W}\dot{Q}'$ and $\ddot{W} \dot{Q}$.
We introduce a Lagrange multiplier $\chi$ that helps us to understand the presence and behavior of propagating DOFs. Then, the Lagrangian equivalent to $L$ can be written as $L-L_K+\tilde{L}_K$ where 
\be
\tilde{L}_K=\frac{3\alpha M_{\rm pl}^2 
\sqrt{h} [2f'h r +f(2-2h-rh')]l (l+1)}
{2f^{5/2} r^2}
\left[ 2\chi \left( \ddot{W} -{\dot Q}' 
+\frac{2\dot{Q}}{r} \right)-\chi^2
\right]\,.
\label{cLK}
\ee
Indeed, the variation of (\ref{cLK}) with respect to $\chi$ leads to $\chi=\ddot{W} -{\dot Q}'+2\dot{Q}/r$, so that (\ref{cLK}) reduces to (\ref{LK}). 
We note that the gauge-invariant field $\chi=\ddot{W} -{\dot Q}'+2\dot{Q}/r$ corresponds to a time derivative of the dynamical perturbation $\tilde{\chi}=\dot{W}-Q'+2Q/r$ in GR \cite{Regge:1957td,Kobayashi:2012kh,Kase:2021mix} (up to a free function of $r$).
Integrating the Lagrangian $\chi \ddot{W}$ by parts in Eq.~(\ref{cLK}), 
we obtain the product $-\dot{W} \dot{\chi}$ between the two first derivatives.
As we will see below, there are three propagating DOFs $\vec{{\cal X}}=(W,\chi,Q)$ in this system. 
In other words, we generally need to give six independent initial conditions to determine the time evolution of $\vec{{\cal X}}$.  

After the integration by parts, the total second-order can be 
expressed in the form ${\cal S}^{(2)}=\int {\rm d}t {\rm d}r\,L$, where
\ba
L &=&
a_1 \dot{W}^2+a_2 \dot{Q}^2
+2 a_3 \dot{W} \dot{\chi}
+a_4 \left( \dot{W}'-Q''+\frac{2Q'}{r} 
\right)^2
+a_5 W'^2+a_6 Q'^2
+a_7 W^2+a_8 \chi^2 +a_9 Q^2
\nonumber \\
& &
+a_{10} W' \dot{Q}
+a_{11} \dot{W} Q'+a_{12}\dot{\chi}Q'
+a_{13} \dot{W}Q
+a_{14} \chi \dot{Q}\,,
\label{Lag}
\ea
with $a_{1}, \cdots ,a_{14}$ being functions of $r$ alone. Varying the Lagrangian (\ref{Lag}) with respect to $W$, $\chi$, and $Q$, we obtain the field equations of motion for these perturbations.
They contain the derivatives up to the fourth order, e.g., $\ddot{W}''$, $\dot{Q}'''$.

In the following, we study the propagation of fast oscillating modes by assuming the solutions in the form
\be
\vec{\cal X}=\vec{\cal X}_0 
e^{i (\omega t-kr)}\,,\qquad 
{\rm with} \qquad 
\vec{\cal X}_0=(W_0, \chi_0, Q_0)\,.
\label{cX}
\ee
Because of staticity, here, the coefficients $\vec{{\cal X}}_0$ are supposed to be functions slowly varying in the $r$-direction, so that in the WKB domain they satisfy for instance the condition $|{\vec{{\cal X}}}_0'| \ll |k\vec{{\cal X}}_0|$. For the same reason, we will also suppose that $|\omega'|\ll |k\omega|\simeq|\omega^2|$. As a consequence, we will also consider the wavenumber $k$ and the multipole $l$ in the ranges $k r_h \gg 1$ and $l \gg 1$.

Then, each coefficient in Eq.~(\ref{Lag}) 
has the following $l$ dependence:
\ba
& &
a_1=b_1 l^4\,,\quad
a_2=b_2 l^4\,,\quad
a_3=b_3 l^2\,,\quad
a_4=b_4 l^2\,,\quad
a_5=b_5 l^4\,,\quad
a_6=b_6 l^4\,,\quad
a_7=b_7 l^6\,,\quad
a_8=b_8 l^2\,,\nonumber \\
& &
a_9=b_9 l^6\,,\quad
a_{10}=b_{10} l^4\,,\quad
a_{11}=b_{11} l^4\,,\quad
a_{12}=b_{12} l^2\,,\quad
a_{13}=b_{13} l^4\,,\quad
a_{14}=b_{14} l^2\,,
\ea
where 
\ba
& &
b_1=\frac{3\alpha \Mpl^2 h^{1/2} 
[fr(2f''hr + f'h'r - 6f'h)+2f^2(h'r 
+ 2h - 2)-f'^2 hr^2]}{4f^{5/2}r^4}\,,
\nonumber \\
& &
b_2=\frac{3\alpha\Mpl^2 
[(2f''f - f'^2)hr + h'f(f'r - 2f)]}
{2f^{7/2}h^{1/2}r^3}\,,
\nonumber \\
& &
b_3=\frac{r^2}{f}b_1-\frac{hr^2}{2}b_2
+\frac{b_0}{2fh}\,,\qquad
b_4=fh b_3+\frac{3}{2}b_0\,,\qquad
b_5=f^2 h^2 b_2-\frac{f}{r^2}b_0\,,\qquad
b_6=b_1+\frac{2}{h r^2}b_0\,,
\nonumber \\
& &
b_7=-\frac{b_5}{2h r^2}\,,\qquad
b_8=b_3\,,\qquad
b_9=\frac{f}{2r^2}b_2\,,\qquad
b_{10}=-\frac{3\alpha\Mpl^2\sqrt{h}(3f' h+ h'f)}
{f^{3/2} r^3}\,,
\nonumber \\
& &
b_{11}=-2b_1-2f h b_2-b_{10}-\frac{b_0}{hr^2}\,,
\qquad 
b_{12}=-2b_3\,,
\ea
with 
\be
b_0=\frac{3\alpha\Mpl^2 h^{3/2}(f'h -h'f)}
{f^{3/2} r}\,.
\ee
We do not show the explicit forms of $b_{13}$ 
and $b_{14}$, as they are not needed in the 
following discussion.

Exploiting the approximation of large values of 
$\omega$, $k$, and $l$ in the field equations 
of motion, we ignore terms proportional to 
$i \omega$, $ik$ and $l$ relative to 
those proportional to $\omega^2$, $k^2$ and $l^2$.
Then, the perturbation equations of motion 
following from the Lagrangian (\ref{Lag}) can be 
expressed in the form 
\be
{\bm A} \vec{{\cal X}}_0^{\rm T}=0\,,
\label{Aeq}
\ee
where ${\bm A}$ is the $3 \times 3$ matrix whose 
components are given by 
\ba
& &
A_{11}=2l^2 \left[(b_4 k^2+b_1 l^2)\omega^2 
+ b_5 k^2 l^2 + b_7 l^4 \right]\,,\qquad
A_{22}=2l^2 b_8\,,\qquad
A_{33}=2 l^2 \left(b_2 l^2 \omega^2 +b_4 k^4 
+ b_6 k^2 l^2 + b_9 l^4 \right)\,,\nonumber\\
& &
A_{12}=A_{21}=2l^2 b_3 \omega^2\,,\qquad
A_{13}=A_{31}=l^2(2b_4k^3-b_{10}k l^2-b_{11}k l^2)
\omega\,,\qquad 
A_{23}=A_{32}=-l^2 b_{12}k\,\omega\,.
\ea
\subsection{Ghost instability}

To derive the no-ghost conditions, we pick up 
the matrix components in ${\bm A}$ containing terms 
proportional to $\omega^2$, i.e., 
\be
{\bm K}=2\omega^2 \left(\begin{array}{ccc}
K_{11} & K_{12} & 0\\
K_{12} & 0 & 0\\
0 & 0 & K_{33}
\end{array}\right)\,,
\ee
where 
\be
K_{11}=l^2 (b_4 k^2+b_1 l^2)\,,\qquad 
K_{12}=l^2 b_3\,,\qquad 
K_{33}=l^4 b_2\,.
\ee
The absence of ghosts requires the following three 
conditions
\be
K_{11}>0\,,\qquad -K_{12}^2>0\,,\qquad 
-K_{12}^2 K_{33}>0\,.
\label{noghost}
\ee
The explicit form of $K_{12}$ is given by 
\be
K_{12}=-\frac{3\alpha \Mpl^2 h^{1/2} 
[r(2f'h -h' f)+2f(1-h)]}{2 f^{5/2}r^2} l^2\,.
\ee
So long as $\alpha \neq 0$, the second condition 
of Eq.~(\ref{noghost}) is always violated and hence 
we have at least one ghost mode. 
The existence of the dynamical perturbation 
$\chi$, which is kinetically coupled to $W$, 
gives rise to the ghost DOF.
This result generally holds without restricting the 
background BH solution to the form (\ref{fh2}), 
as we have not assumed the condition $f=h$.

It is important to understand how many ghost DOFs 
are generated in ECG, at least for the vacuum case. 
A simple way to see this is to make the field redefinitions: $W=W_2-K_{12}\,{\chi_2}/K_{11}$, $\chi=\chi_2$, and $Q=Q_2$. The Jacobian of the transformation always has 
a determinant equal to unity, so that the transformation is well defined. On using these new fields, the kinetic matrix becomes diagonal with elements $K_{11}$, $-K_{12}^2/K_{11}$, and $K_{33}$. For the BH solution (\ref{fh2}),  
which is approximately valid for $r\gg r_h$, we find that
\be
K_{33} \simeq -\frac{9\alpha \Mpl^2 r_h l^4}{r^5} \,,\qquad\frac{K_{33}}{K_{11}} \simeq \frac{2l^2}{k^2r^2}>0\,.
\ee
This shows that, at least for the approximate BH solution, $K_{11}$ and $K_{33}$ share the same sign. 
Therefore, in the range of validity of this solution, 
we would have one ghost only if $\alpha<0$, whereas we would have two ghost DOFs if instead $\alpha>0$. This calculation also shows that the kinetic term of, e.g.,\ the mode $Q$, tends to vanish as $r_h/r\to0$. This limit then needs to be taken with care.

\subsection{Radial propagation}

The propagation speeds $c_r={\rm d}r_*/{\rm d}\tau$ 
along the radial direction, where 
$r_*=\int {\rm d}r/\sqrt{h}$ is the rescaled radial 
coordinate and $\tau=\int \sqrt{f}\,{\rm d}t$ is 
the proper time, are expressed as 
$c_r=(fh)^{-1/2}(\partial\omega/\partial k)$. 
We compute the eigenvalue of the matrix ${\bm A}$ 
and expand it with respect to the large momentum $k$. 
This amounts to considering the modes in the range 
$k r_h \gg l \gg 1$.
The leading-order $k^8$ dependent term vanishes on 
account of the relation $b_{12}=-2b_3$, 
so the dominant terms in ${\rm det}\,{\bm A}$ 
are those proportional to $k^6$. 
Then, we obtain the following three solutions to 
the squared radial propagation speeds:
\ba
c_{r1}^2 &=&
\frac{4b_4 b_8 f^4 r^4}
{9 \alpha^2 \Mpl^4 h^2
[(h'r + 2h - 2)f - 2f' hr]^2}
=1+\frac{3r(f'h-h'f)}
{(h'r + 2h - 2)f - 2f' hr}\,,\label{cr1}\\
c_{r2}^2 &=&
\frac{-(b_1+b_6+b_{10}+b_{11})
+\sqrt{{\cal D}_1}}{2b_2 fh}=1\,,\\
c_{r3}^2 &=&
\frac{-(b_1+b_6+b_{10}+b_{11})
-\sqrt{{\cal D}_1}}{2b_2 fh}
=1-\frac{2f (f'h-h'f)}
{2f''f hr -f'r(h f'-fh')
-2h' f^2}\,,
\label{cr3}
\ea
where
\ba
{\cal D}_1=b_1^2+2b_1 b_6-4 b_2 b_5+b_6^2
+b_{10}^2+2(b_1+b_6+b_{11})b_{10}
+2(b_1+b_6)b_{11}+b_{11}^2
=\frac{9\alpha^2 \Mpl^4 h (f'h-h'f)^2}{f^3 r^6}\,.
\ea
For the BH solution (\ref{fh2}) we have $f=h$, in which case 
all the three squared propagation speeds 
(\ref{cr1})-(\ref{cr3}) reduce to 1.

\subsection{Angular instability}

The propagation speeds $c_{\Omega}
=r {\rm d}\theta/{\rm d}\tau$ 
along the angular direction are expressed as 
$c_{\Omega}=r\omega/(\sqrt{f} l)$. 
Expanding ${\rm det}\,{\bm A}$ with respect to 
large $l$ in the range $l \gg k r_h \gg 1$, 
the leading-order terms, which are 
proportional to $l^{14}$, give the following 
squared angular propagation speeds: 
\ba
c_{\Omega 1}^2 &=& 
\frac{r^2(b_1 b_8+\sqrt{{\cal D}_2})}{2b_3^2 f}=1\,,
\label{cO1}\\
c_{\Omega 2}^2 &=& 
\frac{r^2(b_1 b_8-\sqrt{{\cal D}_2})}{2b_3^2 f}
=\frac{[2f'' f h r -f'(f' hr - h'fr + 2fh)]r}
{2f [r (h'f-2f'h) + 2f(h - 1)]}\,,\label{cO2}\\
c_{\Omega 3}^2 &=& -\frac{b_9 r^2}{b_2 f}
=-\frac{1}{2}\,,
\label{cO3}
\ea
where
\ba
\hspace{-0.3cm}
{\cal D}_2 &=&
b_8 (b_1^2 b_8+4 b_3^2 b_7) \nonumber \\
\hspace{-0.3cm}
&=& 
\frac{81 \alpha^4 \Mpl^8 h^2 
[r(2f'h - h'f) - 2f(h - 1)]^2
[2f'' f h r^2-f' r(f'hr -h'fr - 2fh) 
- 2f^2(h'r + 2h - 2)]^2}
{64f^{10}r^{12}}\,.
\ea
Using the BH solution (\ref{fh2}) in the regime 
$r \gg r_h$, the second propagation speed squared 
(\ref{cO2}) has the following behavior
\be
c_{\Omega 2}^2=1-\frac{720 \alpha r_h}{r^5}
+{\cal O}(r^{-10})\,,
\ee
which quickly approaches 1 at large distances.

The third propagation speed squared (\ref{cO3}) is negative, so there is Laplacian instability along the angular direction. 
This instability arises from the unhealthy propagation of the perturbation $Q$. 
In the region slightly away from the horizon, the typical 
time for the instability to occur is of order 
$T\simeq \mathcal{O}(1)/|\omega| \simeq 
r_h/(\sqrt{f}l)\ll r_h$.
We also note that the condition $f=h$ was not used for the derivation of $c_{\Omega 3}^2$.
Due to this angular instability besides the appearance of a ghost, the SSS BH solutions with unsuppressed cubic-order terms are not 
stable.

\subsection{Instability versus EFT mass breaking limit}
\label{EFTsec}

In what follows, we study the regime of validity of the BH solution and conditions under which our previous analysis can be trusted, especially regarding the presence of instabilities. Indeed, we know that one or two ghosts are present and that the field $Q$ acquires a negative propagation speed squared along the angular direction. 
The last instability is purely (and already) classical, and as such, probably the most serious one. 
We would like to see here whether this instability for the field $Q$ occurs anywhere in the spatial slicing with $l$ in the range $l\gg k r_h$. 
In order to study this, we consider the partial differential equations, focusing on the unstable mode, which corresponds to the mass term of the perturbation $Q$. 

Focusing on the term which makes $Q$ unstable and also 
looking for the case $r \gg r_h$, we find that
\be
\ddot Q \simeq 
\left( \frac{l^2}{2r^2} - \frac{r_h^2}{36\alpha}\,\frac{r^3}{r_h^3}\right)Q+\dots
\label{eq:mass_Q}
\ee
where the dots stand for other terms which are irrelevant 
to the following discussion. 
For example, in Eq.~(\ref{eq:mass_Q}), 
we also have a term of the form
\be
\frac{r^5}{36l^2\alpha r_h}\,Q''\,,
\ee
but in the following, we will show that, for the Laplacian instability to occur, we need to be in a region of 
the spatial geometry (and parameter space) 
for which $l^2\gg r^5/(18|\alpha|r_h)$. 
This sets the coefficient of $Q''$ very small. 
Furthermore, since we are considering the case in which 
the angular instability occurs, we also need to assume 
that the angular derivatives overcome the $r$-derivatives, 
that is, $l^2 |Q|/r^2 \gg |Q''|$. 
For the same reason, the coefficients in the terms 
$-r^5{\dot W'}/(36l^2\alpha r_h)$ and 
$-r^4{\dot W}/(18l^2\alpha r_h)$ are suppressed 
and can be thought to be subdominant, also because 
the field $W$ does not exhibit instabilities along 
the angular direction 
(as it propagates with a speed approximately equal to unity).

In any case, from Eq.~(\ref{eq:mass_Q}), we can define an effective squared mass, the self-coupling term independent of 
$l$ (or $k$), for the variable $Q$ as follows
\be
\mu^2_{Q}=\frac{r_h^2}{36|\alpha|}\,\frac{r^3}{r_h^3}\,,
\ee
where we suppose for the moment that $\alpha>0$. 
We see that a negative value of $\alpha$ would make the instability even worse, but our goal here is not trying to solve the partial differential equations, even approximately, for $r\to\infty$. This is because, in fact, it is clear that the term proportional to $l^2$ in Eq.~(\ref{eq:mass_Q}) 
decreases as $r$ increases, whereas the opposite happens for $\mu^2_{Q}$. Evidently, this increase in the mass of the 
perturbation $Q$ cannot approach infinity. 
In fact, we would expect that the theory reaches 
an EFT cutoff so that it needs to be ultraviolet (UV) completed. In other words, this behavior implies that the asymptotically flat limit for the theory 
needs a UV completion. 
Let us call this cutoff mass scale $M \lesssim \Mpl$. 
Then, in this case, the instability really occurs only if
\be
\frac{r_h^2}{36|\alpha|}\,\frac{r^3}{r_h^3} \ll M^2\,.
\label{Mlimit}
\ee
This implies that, if the instability occurs at about 
$r=\beta r_h$, then we require that 
\be
\Mpl^2 \gtrsim M^2 \gg \beta^3 \frac{r_h^2}{36|\alpha|}\,
\qquad {\rm or} \qquad
|\alpha|\gg \frac{\beta^3}{36(4\pi)^2}\,\frac{M_{S}^2}{\Mpl^4 M^2}\,,\label{eq:EFT_alpha}
\ee
where $M_S$ is the BH mass related to $r_h$ 
as $r_h=M_S/(4\pi \Mpl^2)$. 
If the inequality (\ref{eq:EFT_alpha}) does not hold, 
we should expect that the solution is outside 
the EFT domain. In this case, it should either be replaced by another solution (maybe time-dependent) or instead, the spherically symmetric description of gravity in this theory already requires an appropriate UV completion.

At the same time, we require that the BH solution 
(\ref{fh2}), if it is inside the EFT domain, should not be 
so different from the standard Schwarzschild solution. 
This statement holds true if
\be
\frac{54|\alpha| r_h^2}{r^6}\ll \frac{r_h}{r}\,,
\ee
in particular for $r=\beta r_h$. 
This condition leads to
\be
|\alpha|\ll \frac{\beta^5}
{54 (4\pi)^4}\, 
\frac{M_S^4}{\Mpl^8}\,.
\ee
Putting these two requirements on $\alpha$ together, 
we find
\be
\alpha_{\rm min}^{r=\beta r_h}\equiv
\frac{\beta^3}{36 (4\pi)^2}\,\frac{M_{S}^2}{\Mpl^4 M^2}
\ll |\alpha| \ll \alpha_{\rm max}^{r=\beta r_h}\equiv
\frac{\beta^5}{54(4\pi)^4}\,\frac{M_S^4}{\Mpl^8}\,.
\label{eq:cond_alpha}
\ee
This condition makes sense only if $\alpha_{\rm min}^{r=\beta r_h}\ll \alpha_{\rm max}^{r=\beta r_h}$, that is $M\gg (\Mpl/M_S)\Mpl$, i.e.\ $M\gg5.3\times10^{-21}$~GeV for a solar mass BH. A larger BH would make this bound weaker. Hence, in the following, we will consider the range 
$(\Mpl/M_S)\Mpl\ll M \lesssim \Mpl$. 
It should be noted that this relation depends on the BH mass. 
Expressing the dimensionful coupling $\alpha$ in terms of 
a length scale $\ell_{\alpha}$ as $\alpha=\ell_{\alpha}^4$ and saturating the cutoff mass scale to $M\simeq\Mpl$ in order to be able to discuss a more concrete example, the inequality (\ref{eq:cond_alpha}) translates to 
\be
\frac{\beta^{3/4}}{2\sqrt{6\pi}} \left( \frac{M_S}{\Mpl} \right)^{1/2} 
\ell_{\rm Pl} \ll |\ell_{\alpha}| \ll \frac{\beta^{5/4}}{4\pi(54)^{1/4}}
\frac{M_S}{\Mpl} \ell_{\rm Pl}\,,
\label{eq:cond_alpha2}
\ee
where $\ell_{\rm Pl}=1/\Mpl=8.1 \times 10^{-35}$~m is the reduced Planck length. 
If we consider a BH with $M_S=M_{\odot}=2.0 \times 10^{30}$~kg and 
$\beta=2$, for instance, the inequality (\ref{eq:cond_alpha2}) corresponds 
to $3.4 \times 10^{-16}~{\rm m} \ll |\ell_{\alpha}| \ll 2.6 \times 10^{3}$~m. 
For $|\ell_{\alpha}|$ close to the upper limit of (\ref{eq:cond_alpha2}), 
the cubic curvature terms can give 
rise to appreciable deviation from BHs in GR in the vicinity of the horizon 
($r_h \simeq 3 \times 10^3$~m). 
For increasing $M_S$ and $\beta$, both the upper and lower limits of 
$|\ell_{\alpha}|$ tend to be larger.

Now, if the relation (\ref{eq:cond_alpha}) or (\ref{eq:cond_alpha2}) holds, then we can trust the solution as a valid EFT, but it would be unstable, because of the negative angular squared speed of propagation. Indeed, this instability would take place if
\be
\frac{l^2}{2r^2}\gg \mu_Q^2\,,\qquad{\rm or}
\qquad
l^2\gg \frac{r^5}{18|\alpha| r_h} \gg 1\,.
\ee
Assuming also that $l^2/(2r^2)\ll M^2\lesssim \Mpl^2$, 
we find
\be
3\,\frac{\alpha_{\rm max}^{r=\beta r_h}}{|\alpha|}\ll l^2 \ll \frac{\beta^2}{8\pi^2}\frac{M_S^2\,M^2}{\Mpl^4}\,.
\ee
These last inequalities are compatible with the conditions (\ref{eq:cond_alpha}).

We do not find it interesting to discuss the behavior 
of the BH solution in the limit $r\to\infty$, as in this case, 
we soon hit the EFT breaking scale, and hence 
the analysis would lose its validity.

After the initial submission of this paper, we noticed that Bueno {\it et al.}
put a paper on the arXiv claiming that the BH instability can be avoided 
within the EFT regime of ECG \cite{Bueno:2023jtc}. 
First of all, we would like to stress that, unlike the discussion given 
above, the EFT approach taken in Ref.~\cite{Bueno:2023jtc} means that 
the cubic Lagrangian $\ell_\alpha^4 {\cal P}$ is always suppressed 
relative to the Einstein-Hilbert term. In this case, the cutoff mass scale $M$ for the validity of the theory is related to the cubic coupling constant 
$\ell_\alpha^4$ as $M=1/\ell_\alpha$. As also the authors admit, the difference in our approach consists of relaxing this bound and considering instead general values for the cutoff $M$. In this sense, our approach is more general, but it is true, on the other hand, that our bound does not apply to all the parameter spaces 
in ECG theories. For the choice $M=1/\ell_\alpha$, 
the condition (\ref{Mlimit}) 
for the occurrence of Laplacian instability translates to 
$r^3 \ll 36 \ell_\alpha^2 r_h$. The EFT approach requires that 
the corrections induced by the cubic Lagrangian are small outside the 
BH horizon. Since $\ell_{\alpha} \ll r_h$ in this case, the inequality 
(\ref{Mlimit}) holds in the regime $r \ll r_h$, i.e., deep inside the horizon.

As we already mentioned at the beginning of this section, unlike Ref.~\cite{Bueno:2023jtc}, we have not dealt with the ECG as an EFT where the cubic Lagrangian $\ell_\alpha^4 {\cal P}$ is always suppressed relative to $R$. In this case, the cutoff mass scale $M$ in Eq.~(\ref{Mlimit}) is not necessarily restricted to taking values of the order of $1/\ell_\alpha$. 
Then, the Laplacian instability is present outside the horizon ($\beta \geq 1$) for the coupling constant $\ell_{\alpha}$ in the wide range of (\ref{eq:cond_alpha2}). 
In other words, whenever the cubic Lagrangian 
starts to be comparable to the Ricci scalar, the problem of Laplacian instability emerges besides the ghost and strong coupling problems.
Since we always need to be in the EFT domain with 
$|\ell_\alpha^4 {\cal P}| \ll |R|$ to avoid this pathological behavior, 
it is not possible to deal with any nonperturbative phenomenon in the vicinity of the horizon \cite{Jimenez:2023esa}. 
Whenever nonperturbative effects come into play in BH physics, we hit the aforementioned problems due to the breakdown of EFT.

\section{Conclusions}
\label{consec}

ECG is a cubic-order gravitational theory constructed to share the massless graviton spectrum similar to that in GR on a maximally symmetric background. 
Since the Lagrangian ${\cal P}$ is beyond the domain of Lovelock theories, the field equations of motion contain derivatives higher than second order in general, curved geometries. At the same time, this suggests that there should be additional propagating DOFs to those in GR for the spacetime different from the maximally symmetric background. On the SSS background given by the line element (\ref{background}), we studied the propagation of odd-parity perturbations and the resulting linear stability conditions of propagating DOFs. 
In this procedure, we did not deal with the ECG as a rather trivial EFT where the cubic Lagrangian is always strongly suppressed relative to the Einstein-Hilbert term.

At the background level, the metric components of SSS vacuum solutions in ECG obey the third-order single differential 
Eq.~(\ref{difeq}) with $f=h$. 
At large distances away from the horizon, the SSS BH solution is approximately given by Eq.~(\ref{fh2}). 
On using the large-distance expansion (\ref{fhex}) of metrics without imposing the condition $f=h$, we also obtain the same result as Eq.~(\ref{fh2}). 
This shows the universality of SSS BH solutions with $f=h$.

In Sec.~\ref{persec}, we expanded the action (\ref{action2}) 
in ECG up to quadratic order in odd-parity perturbations 
to see the propagation of dynamical DOFs. 
There are some higher-order derivatives appearing as the form (\ref{LK}) in the action. 
By introducing a gauge-invariant Lagrange multiplier $\chi=\ddot W -{\dot Q}' +2\dot{Q}/r$, one can express Eq.~(\ref{LK}) as the equivalent Lagrangian (\ref{cLK}).
Then, the total Lagrangian is expressed by the form (\ref{Lag}), which consists of three dynamical perturbations $\vec{{\cal X}}=(W,\chi,Q)$.
In comparison to GR, which contains only one dynamical perturbation $\tilde{\chi}=\dot{W}-Q'+2Q/r$ 
in the odd-parity sector, there are two more 
propagating DOFs in ECG.

For high frequencies and large radial and angular momenta, we derived the perturbation equations in the form (\ref{Aeq}) by assuming the solution as the WKB form (\ref{cX}). 
The ghosts are absent under the three conditions (\ref{noghost}), but the second condition is always violated for $\alpha \neq 0$. 
This is the outcome of a kinetic mixing of $\dot{\chi}$ and $\dot{W}$ without the $\dot{\chi}^2$ term. 
We showed that all the radial propagation speeds reduce to 1 for the BH solution (\ref{fh2}). 
Along the angular direction, the squared propagation speed of the perturbation $Q$ is negative ($c_{\Omega 3}^2=-1/2$) 
for the coupling constant $\alpha$ in the range (\ref{eq:cond_alpha}). 
Thus, the SSS BH in ECG with unsuppressed higher-order curvature terms
is excluded by both ghost and Laplacian instabilities. 

The results found in this paper are analogous to what happens in the anisotropic cosmological background in CECG \cite{Pookkillath:2020iqq}. 
In this case, there are also three propagating DOFs, 
with the appearance of ghosts and tachyonic instabilities on the quasi-de Sitter background with small anisotropies. 
We also note that a strong coupling problem arises for both ECG and CECG 
due to the degeneracy of propagating DOFs around the maximally symmetric background \cite{BeltranJimenez:2020lee,Jimenez:2023esa}. 
These pathologies may be avoided by restricting ourselves to the EFT domain in which the cubic Lagrangians are 
strongly suppressed relative to the Einstein-Hilbert term, but the problems of ghosts, instabilities, and strong couplings manifest themselves in the regime where the cubic curvature terms become comparable to $R$.
These results suggest that going beyond the Lovelock domain with unsuppressed cubic curvature terms can cause problems in the spacetime geometry different from the maximally symmetric background. 
It will be of interest to explore further whether the similar property persists or not for higher-order gravitational theories containing quadratic and quartic curvature terms.

\section*{Acknowledgements}

We thank Jose Beltr\'an Jim\'enez and Alejandro Jim\'enez-Cano for useful correspondence. The work of ADF was supported by the Japan Society for the Promotion of Science Grants-in-Aid for Scientific Research No.\ 20K03969 and by grant PID2020-118159GB-C41 funded by MCIN/AEI/10.13039/501100011033. 
ST is supported by the Grant-in-Aid for Scientific Research 
Fund of the JSPS No.~22K03642 and Waseda University 
Special Research Project No.~2023C-473.


\bibliographystyle{mybibstyle}
\bibliography{bib}

\begin{thebibliography}{56}%
\makeatletter
\providecommand \@ifxundefined [1]{%
 \@ifx{#1\undefined}
}%
\providecommand \@ifnum [1]{%
 \ifnum #1\expandafter \@firstoftwo
 \else \expandafter \@secondoftwo
 \fi
}%
\providecommand \@ifx [1]{%
 \ifx #1\expandafter \@firstoftwo
 \else \expandafter \@secondoftwo
 \fi
}%
\providecommand \natexlab [1]{#1}%
\providecommand \enquote  [1]{``#1''}%
\providecommand \bibnamefont  [1]{#1}%
\providecommand \bibfnamefont [1]{#1}%
\providecommand \citenamefont [1]{#1}%
\providecommand \href@noop [0]{\@secondoftwo}%
\providecommand \href [0]{\begingroup \@sanitize@url \@href}%
\providecommand \@href[1]{\@@startlink{#1}\@@href}%
\providecommand \@@href[1]{\endgroup#1\@@endlink}%
\providecommand \@sanitize@url [0]{\catcode `\\12\catcode `\$12\catcode
  `\&12\catcode `\#12\catcode `\^12\catcode `\_12\catcode `\%12\relax}%
\providecommand \@@startlink[1]{}%
\providecommand \@@endlink[0]{}%
\providecommand \url  [0]{\begingroup\@sanitize@url \@url }%
\providecommand \@url [1]{\endgroup\@href {#1}{\urlprefix }}%
\providecommand \urlprefix  [0]{URL }%
\providecommand \Eprint [0]{\href }%
\providecommand \doibase [0]{http://dx.doi.org/}%
\providecommand \selectlanguage [0]{\@gobble}%
\providecommand \bibinfo  [0]{\@secondoftwo}%
\providecommand \bibfield  [0]{\@secondoftwo}%
\providecommand \translation [1]{[#1]}%
\providecommand \BibitemOpen [0]{}%
\providecommand \bibitemStop [0]{}%
\providecommand \bibitemNoStop [0]{.\EOS\space}%
\providecommand \EOS [0]{\spacefactor3000\relax}%
\providecommand \BibitemShut  [1]{\csname bibitem#1\endcsname}%
\let\auto@bib@innerbib\@empty
\bibitem [{\citenamefont {Hoyle}\ \emph {et~al.}(2001)\citenamefont {Hoyle},
  \citenamefont {Schmidt}, \citenamefont {Heckel}, \citenamefont {Adelberger},
  \citenamefont {Gundlach}, \citenamefont {Kapner},\ and\ \citenamefont
  {Swanson}}]{Hoyle:2000cv}%
  \BibitemOpen
  \bibfield  {author} {\bibinfo {author} {\bibfnamefont {C.~D.}\ \bibnamefont
  {Hoyle}}, \bibinfo {author} {\bibfnamefont {U.}~\bibnamefont {Schmidt}},
  \bibinfo {author} {\bibfnamefont {B.~R.}\ \bibnamefont {Heckel}}, \bibinfo
  {author} {\bibfnamefont {E.~G.}\ \bibnamefont {Adelberger}}, \bibinfo
  {author} {\bibfnamefont {J.~H.}\ \bibnamefont {Gundlach}}, \bibinfo {author}
  {\bibfnamefont {D.~J.}\ \bibnamefont {Kapner}},  and \bibinfo {author}
  {\bibfnamefont {H.~E.}\ \bibnamefont {Swanson}},\ }\href {\doibase
  10.1103/PhysRevLett.86.1418} {\bibfield  {journal} {\bibinfo  {journal}
  {\emph {Phys. Rev. Lett.}}\ }\textbf {\bibinfo {volume} {86}},\ \bibinfo
  {pages} {1418} (\bibinfo {year} {2001})},\ \Eprint
  {http://arxiv.org/abs/hep-ph/0011014} {arXiv:hep-ph/0011014} \BibitemShut
  {NoStop}%
\bibitem [{\citenamefont {Adelberger}\ \emph {et~al.}(2003)\citenamefont
  {Adelberger}, \citenamefont {Heckel},\ and\ \citenamefont
  {Nelson}}]{Adelberger:2003zx}%
  \BibitemOpen
  \bibfield  {author} {\bibinfo {author} {\bibfnamefont {E.~G.}\ \bibnamefont
  {Adelberger}}, \bibinfo {author} {\bibfnamefont {B.~R.}\ \bibnamefont
  {Heckel}},  and \bibinfo {author} {\bibfnamefont {A.~E.}\ \bibnamefont
  {Nelson}},\ }\href {\doibase 10.1146/annurev.nucl.53.041002.110503}
  {\bibfield  {journal} {\bibinfo  {journal} {\emph {Ann. Rev. Nucl. Part.
  Sci.}}\ }\textbf {\bibinfo {volume} {53}},\ \bibinfo {pages} {77} (\bibinfo
  {year} {2003})},\ \Eprint {http://arxiv.org/abs/hep-ph/0307284}
  {arXiv:hep-ph/0307284} \BibitemShut {NoStop}%
\bibitem [{\citenamefont {De~Felice}\ and\ \citenamefont
  {Tsujikawa}(2010)}]{DeFelice:2010aj}%
  \BibitemOpen
  \bibfield  {author} {\bibinfo {author} {\bibfnamefont {A.}~\bibnamefont
  {De~Felice}} and \bibinfo {author} {\bibfnamefont {S.}~\bibnamefont
  {Tsujikawa}},\ }\href {\doibase 10.12942/lrr-2010-3} {\bibfield  {journal}
  {\bibinfo  {journal} {\emph {Living Rev. Rel.}}\ }\textbf {\bibinfo {volume}
  {13}},\ \bibinfo {pages} {3} (\bibinfo {year} {2010})},\ \Eprint
  {http://arxiv.org/abs/1002.4928} {arXiv:1002.4928 [gr-qc]} \BibitemShut
  {NoStop}%
\bibitem [{\citenamefont {Clifton}\ \emph {et~al.}(2012)\citenamefont
  {Clifton}, \citenamefont {Ferreira}, \citenamefont {Padilla},\ and\
  \citenamefont {Skordis}}]{Clifton:2011jh}%
  \BibitemOpen
  \bibfield  {author} {\bibinfo {author} {\bibfnamefont {T.}~\bibnamefont
  {Clifton}}, \bibinfo {author} {\bibfnamefont {P.~G.}\ \bibnamefont
  {Ferreira}}, \bibinfo {author} {\bibfnamefont {A.}~\bibnamefont {Padilla}},
  and \bibinfo {author} {\bibfnamefont {C.}~\bibnamefont {Skordis}},\ }\href
  {\doibase 10.1016/j.physrep.2012.01.001} {\bibfield  {journal} {\bibinfo
  {journal} {\emph {Phys. Rept.}}\ }\textbf {\bibinfo {volume} {513}},\
  \bibinfo {pages} {1} (\bibinfo {year} {2012})},\ \Eprint
  {http://arxiv.org/abs/1106.2476} {arXiv:1106.2476 [astro-ph.CO]} \BibitemShut
  {NoStop}%
\bibitem [{\citenamefont {Will}(2014)}]{Will:2014kxa}%
  \BibitemOpen
  \bibfield  {author} {\bibinfo {author} {\bibfnamefont {C.~M.}\ \bibnamefont
  {Will}},\ }\href {\doibase 10.12942/lrr-2014-4} {\bibfield  {journal}
  {\bibinfo  {journal} {\emph {Living Rev. Rel.}}\ }\textbf {\bibinfo {volume}
  {17}},\ \bibinfo {pages} {4} (\bibinfo {year} {2014})},\ \Eprint
  {http://arxiv.org/abs/1403.7377} {arXiv:1403.7377 [gr-qc]} \BibitemShut
  {NoStop}%
\bibitem [{\citenamefont {Stelle}(1977)}]{Stelle:1976gc}%
  \BibitemOpen
  \bibfield  {author} {\bibinfo {author} {\bibfnamefont {K.~S.}\ \bibnamefont
  {Stelle}},\ }\href {\doibase 10.1103/PhysRevD.16.953} {\bibfield  {journal}
  {\bibinfo  {journal} {\emph {Phys. Rev. D}}\ }\textbf {\bibinfo {volume}
  {16}},\ \bibinfo {pages} {953} (\bibinfo {year} {1977})}\BibitemShut
  {NoStop}%
\bibitem [{\citenamefont {Zwiebach}(1985)}]{Zwiebach:1985uq}%
  \BibitemOpen
  \bibfield  {author} {\bibinfo {author} {\bibfnamefont {B.}~\bibnamefont
  {Zwiebach}},\ }\href {\doibase 10.1016/0370-2693(85)91616-8} {\bibfield
  {journal} {\bibinfo  {journal} {\emph {Phys. Lett. B}}\ }\textbf {\bibinfo
  {volume} {156}},\ \bibinfo {pages} {315} (\bibinfo {year}
  {1985})}\BibitemShut {NoStop}%
\bibitem [{\citenamefont {Gross}\ and\ \citenamefont
  {Sloan}(1987)}]{Gross:1986mw}%
  \BibitemOpen
  \bibfield  {author} {\bibinfo {author} {\bibfnamefont {D.~J.}\ \bibnamefont
  {Gross}} and \bibinfo {author} {\bibfnamefont {J.~H.}\ \bibnamefont
  {Sloan}},\ }\href {\doibase 10.1016/0550-3213(87)90465-2} {\bibfield
  {journal} {\bibinfo  {journal} {\emph {Nucl. Phys. B}}\ }\textbf {\bibinfo
  {volume} {291}},\ \bibinfo {pages} {41} (\bibinfo {year} {1987})}\BibitemShut
  {NoStop}%
\bibitem [{\citenamefont {Hofman}(2009)}]{Hofman:2009ug}%
  \BibitemOpen
  \bibfield  {author} {\bibinfo {author} {\bibfnamefont {D.~M.}\ \bibnamefont
  {Hofman}},\ }\href {\doibase 10.1016/j.nuclphysb.2009.08.001} {\bibfield
  {journal} {\bibinfo  {journal} {\emph {Nucl. Phys. B}}\ }\textbf {\bibinfo
  {volume} {823}},\ \bibinfo {pages} {174} (\bibinfo {year} {2009})},\ \Eprint
  {http://arxiv.org/abs/0907.1625} {arXiv:0907.1625 [hep-th]} \BibitemShut
  {NoStop}%
\bibitem [{\citenamefont {Myers}\ and\ \citenamefont
  {Robinson}(2010)}]{Myers:2010ru}%
  \BibitemOpen
  \bibfield  {author} {\bibinfo {author} {\bibfnamefont {R.~C.}\ \bibnamefont
  {Myers}} and \bibinfo {author} {\bibfnamefont {B.}~\bibnamefont {Robinson}},\
  }\href {\doibase 10.1007/JHEP08(2010)067} {\bibfield  {journal} {\bibinfo
  {journal} {\emph {JHEP}}\ }\textbf {\bibinfo {volume} {08}},\ \bibinfo
  {pages} {067} (\bibinfo {year} {2010})},\ \Eprint
  {http://arxiv.org/abs/1003.5357} {arXiv:1003.5357 [gr-qc]} \BibitemShut
  {NoStop}%
\bibitem [{\citenamefont {Myers}\ \emph {et~al.}(2010)\citenamefont {Myers},
  \citenamefont {Paulos},\ and\ \citenamefont {Sinha}}]{Myers:2010jv}%
  \BibitemOpen
  \bibfield  {author} {\bibinfo {author} {\bibfnamefont {R.~C.}\ \bibnamefont
  {Myers}}, \bibinfo {author} {\bibfnamefont {M.~F.}\ \bibnamefont {Paulos}},
  and \bibinfo {author} {\bibfnamefont {A.}~\bibnamefont {Sinha}},\ }\href
  {\doibase 10.1007/JHEP08(2010)035} {\bibfield  {journal} {\bibinfo  {journal}
  {\emph {JHEP}}\ }\textbf {\bibinfo {volume} {08}},\ \bibinfo {pages} {035}
  (\bibinfo {year} {2010})},\ \Eprint {http://arxiv.org/abs/1004.2055}
  {arXiv:1004.2055 [hep-th]} \BibitemShut {NoStop}%
\bibitem [{\citenamefont {Oliva}\ and\ \citenamefont
  {Ray}(2010)}]{Oliva:2010eb}%
  \BibitemOpen
  \bibfield  {author} {\bibinfo {author} {\bibfnamefont {J.}~\bibnamefont
  {Oliva}} and \bibinfo {author} {\bibfnamefont {S.}~\bibnamefont {Ray}},\
  }\href {\doibase 10.1088/0264-9381/27/22/225002} {\bibfield  {journal}
  {\bibinfo  {journal} {\emph {Class. Quant. Grav.}}\ }\textbf {\bibinfo
  {volume} {27}},\ \bibinfo {pages} {225002} (\bibinfo {year} {2010})},\
  \Eprint {http://arxiv.org/abs/1003.4773} {arXiv:1003.4773 [gr-qc]}
  \BibitemShut {NoStop}%
\bibitem [{\citenamefont {Ostrogradsky}(1850)}]{Ostrogradsky:1850fid}%
  \BibitemOpen
  \bibfield  {author} {\bibinfo {author} {\bibfnamefont {M.}~\bibnamefont
  {Ostrogradsky}},\ }\href@noop {} {\bibfield  {journal} {\bibinfo  {journal}
  {\emph {Mem. Acad. St. Petersbourg}}\ }\textbf {\bibinfo {volume} {6}},\
  \bibinfo {pages} {385} (\bibinfo {year} {1850})}\BibitemShut {NoStop}%
\bibitem [{\citenamefont {Woodard}(2015)}]{Woodard:2015zca}%
  \BibitemOpen
  \bibfield  {author} {\bibinfo {author} {\bibfnamefont {R.~P.}\ \bibnamefont
  {Woodard}},\ }\href {\doibase 10.4249/scholarpedia.32243} {\bibfield
  {journal} {\bibinfo  {journal} {\emph {Scholarpedia}}\ }\textbf {\bibinfo
  {volume} {10}},\ \bibinfo {pages} {32243} (\bibinfo {year} {2015})},\ \Eprint
  {http://arxiv.org/abs/1506.02210} {arXiv:1506.02210 [hep-th]} \BibitemShut
  {NoStop}%
\bibitem [{\citenamefont {Lanczos}(1938)}]{Lanczos:1938sf}%
  \BibitemOpen
  \bibfield  {author} {\bibinfo {author} {\bibfnamefont {C.}~\bibnamefont
  {Lanczos}},\ }\href {\doibase 10.2307/1968467} {\bibfield  {journal}
  {\bibinfo  {journal} {\emph {Annals Math.}}\ }\textbf {\bibinfo {volume}
  {39}},\ \bibinfo {pages} {842} (\bibinfo {year} {1938})}\BibitemShut
  {NoStop}%
\bibitem [{\citenamefont {Lovelock}(1971)}]{Lovelock:1971yv}%
  \BibitemOpen
  \bibfield  {author} {\bibinfo {author} {\bibfnamefont {D.}~\bibnamefont
  {Lovelock}},\ }\href {\doibase 10.1063/1.1665613} {\bibfield  {journal}
  {\bibinfo  {journal} {\emph {J. Math. Phys.}}\ }\textbf {\bibinfo {volume}
  {12}},\ \bibinfo {pages} {498} (\bibinfo {year} {1971})}\BibitemShut
  {NoStop}%
\bibitem [{\citenamefont {Metsaev}\ and\ \citenamefont
  {Tseytlin}(1987)}]{Metsaev:1987zx}%
  \BibitemOpen
  \bibfield  {author} {\bibinfo {author} {\bibfnamefont {R.~R.}\ \bibnamefont
  {Metsaev}} and \bibinfo {author} {\bibfnamefont {A.~A.}\ \bibnamefont
  {Tseytlin}},\ }\href {\doibase 10.1016/0550-3213(87)90077-0} {\bibfield
  {journal} {\bibinfo  {journal} {\emph {Nucl. Phys. B}}\ }\textbf {\bibinfo
  {volume} {293}},\ \bibinfo {pages} {385} (\bibinfo {year}
  {1987})}\BibitemShut {NoStop}%
\bibitem [{\citenamefont {Antoniadis}\ \emph {et~al.}(1994)\citenamefont
  {Antoniadis}, \citenamefont {Rizos},\ and\ \citenamefont
  {Tamvakis}}]{Antoniadis:1993jc}%
  \BibitemOpen
  \bibfield  {author} {\bibinfo {author} {\bibfnamefont {I.}~\bibnamefont
  {Antoniadis}}, \bibinfo {author} {\bibfnamefont {J.}~\bibnamefont {Rizos}},
  and \bibinfo {author} {\bibfnamefont {K.}~\bibnamefont {Tamvakis}},\ }\href
  {\doibase 10.1016/0550-3213(94)90120-1} {\bibfield  {journal} {\bibinfo
  {journal} {\emph {Nucl. Phys. B}}\ }\textbf {\bibinfo {volume} {415}},\
  \bibinfo {pages} {497} (\bibinfo {year} {1994})},\ \Eprint
  {http://arxiv.org/abs/hep-th/9305025} {arXiv:hep-th/9305025} \BibitemShut
  {NoStop}%
\bibitem [{\citenamefont {Gasperini}\ \emph {et~al.}(1997)\citenamefont
  {Gasperini}, \citenamefont {Maggiore},\ and\ \citenamefont
  {Veneziano}}]{Gasperini:1996fu}%
  \BibitemOpen
  \bibfield  {author} {\bibinfo {author} {\bibfnamefont {M.}~\bibnamefont
  {Gasperini}}, \bibinfo {author} {\bibfnamefont {M.}~\bibnamefont {Maggiore}},
   and \bibinfo {author} {\bibfnamefont {G.}~\bibnamefont {Veneziano}},\ }\href
  {\doibase 10.1016/S0550-3213(97)00149-1} {\bibfield  {journal} {\bibinfo
  {journal} {\emph {Nucl. Phys. B}}\ }\textbf {\bibinfo {volume} {494}},\
  \bibinfo {pages} {315} (\bibinfo {year} {1997})},\ \Eprint
  {http://arxiv.org/abs/hep-th/9611039} {arXiv:hep-th/9611039} \BibitemShut
  {NoStop}%
\bibitem [{\citenamefont {Kanti}\ \emph {et~al.}(1996)\citenamefont {Kanti},
  \citenamefont {Mavromatos}, \citenamefont {Rizos}, \citenamefont {Tamvakis},\
  and\ \citenamefont {Winstanley}}]{Kanti:1995vq}%
  \BibitemOpen
  \bibfield  {author} {\bibinfo {author} {\bibfnamefont {P.}~\bibnamefont
  {Kanti}}, \bibinfo {author} {\bibfnamefont {N.~E.}\ \bibnamefont
  {Mavromatos}}, \bibinfo {author} {\bibfnamefont {J.}~\bibnamefont {Rizos}},
  \bibinfo {author} {\bibfnamefont {K.}~\bibnamefont {Tamvakis}},  and \bibinfo
  {author} {\bibfnamefont {E.}~\bibnamefont {Winstanley}},\ }\href {\doibase
  10.1103/PhysRevD.54.5049} {\bibfield  {journal} {\bibinfo  {journal} {\emph
  {Phys. Rev. D}}\ }\textbf {\bibinfo {volume} {54}},\ \bibinfo {pages} {5049}
  (\bibinfo {year} {1996})},\ \Eprint {http://arxiv.org/abs/hep-th/9511071}
  {arXiv:hep-th/9511071} \BibitemShut {NoStop}%
\bibitem [{\citenamefont {Torii}\ \emph {et~al.}(1997)\citenamefont {Torii},
  \citenamefont {Yajima},\ and\ \citenamefont {Maeda}}]{Torii:1996yi}%
  \BibitemOpen
  \bibfield  {author} {\bibinfo {author} {\bibfnamefont {T.}~\bibnamefont
  {Torii}}, \bibinfo {author} {\bibfnamefont {H.}~\bibnamefont {Yajima}},  and
  \bibinfo {author} {\bibfnamefont {K.-i.}\ \bibnamefont {Maeda}},\ }\href
  {\doibase 10.1103/PhysRevD.55.739} {\bibfield  {journal} {\bibinfo  {journal}
  {\emph {Phys. Rev. D}}\ }\textbf {\bibinfo {volume} {55}},\ \bibinfo {pages}
  {739} (\bibinfo {year} {1997})},\ \Eprint
  {http://arxiv.org/abs/gr-qc/9606034} {arXiv:gr-qc/9606034} \BibitemShut
  {NoStop}%
\bibitem [{\citenamefont {Kawai}\ \emph {et~al.}(1998)\citenamefont {Kawai},
  \citenamefont {Sakagami},\ and\ \citenamefont {Soda}}]{Kawai:1998ab}%
  \BibitemOpen
  \bibfield  {author} {\bibinfo {author} {\bibfnamefont {S.}~\bibnamefont
  {Kawai}}, \bibinfo {author} {\bibfnamefont {M.-a.}\ \bibnamefont {Sakagami}},
   and \bibinfo {author} {\bibfnamefont {J.}~\bibnamefont {Soda}},\ }\href
  {\doibase 10.1016/S0370-2693(98)00925-3} {\bibfield  {journal} {\bibinfo
  {journal} {\emph {Phys. Lett. B}}\ }\textbf {\bibinfo {volume} {437}},\
  \bibinfo {pages} {284} (\bibinfo {year} {1998})},\ \Eprint
  {http://arxiv.org/abs/gr-qc/9802033} {arXiv:gr-qc/9802033} \BibitemShut
  {NoStop}%
\bibitem [{\citenamefont {Cartier}\ \emph {et~al.}(2000)\citenamefont
  {Cartier}, \citenamefont {Copeland},\ and\ \citenamefont
  {Madden}}]{Cartier:1999vk}%
  \BibitemOpen
  \bibfield  {author} {\bibinfo {author} {\bibfnamefont {C.}~\bibnamefont
  {Cartier}}, \bibinfo {author} {\bibfnamefont {E.~J.}\ \bibnamefont
  {Copeland}},  and \bibinfo {author} {\bibfnamefont {R.}~\bibnamefont
  {Madden}},\ }\href {\doibase 10.1088/1126-6708/2000/01/035} {\bibfield
  {journal} {\bibinfo  {journal} {\emph {JHEP}}\ }\textbf {\bibinfo {volume}
  {01}},\ \bibinfo {pages} {035} (\bibinfo {year} {2000})},\ \Eprint
  {http://arxiv.org/abs/hep-th/9910169} {arXiv:hep-th/9910169} \BibitemShut
  {NoStop}%
\bibitem [{\citenamefont {Tsujikawa}\ \emph {et~al.}(2002)\citenamefont
  {Tsujikawa}, \citenamefont {Brandenberger},\ and\ \citenamefont
  {Finelli}}]{Tsujikawa:2002qc}%
  \BibitemOpen
  \bibfield  {author} {\bibinfo {author} {\bibfnamefont {S.}~\bibnamefont
  {Tsujikawa}}, \bibinfo {author} {\bibfnamefont {R.}~\bibnamefont
  {Brandenberger}},  and \bibinfo {author} {\bibfnamefont {F.}~\bibnamefont
  {Finelli}},\ }\href {\doibase 10.1103/PhysRevD.66.083513} {\bibfield
  {journal} {\bibinfo  {journal} {\emph {Phys. Rev. D}}\ }\textbf {\bibinfo
  {volume} {66}},\ \bibinfo {pages} {083513} (\bibinfo {year} {2002})},\
  \Eprint {http://arxiv.org/abs/hep-th/0207228} {arXiv:hep-th/0207228}
  \BibitemShut {NoStop}%
\bibitem [{\citenamefont {Aoki}\ and\ \citenamefont
  {Tsujikawa}(2023)}]{Aoki:2023jvt}%
  \BibitemOpen
  \bibfield  {author} {\bibinfo {author} {\bibfnamefont {K.}~\bibnamefont
  {Aoki}} and \bibinfo {author} {\bibfnamefont {S.}~\bibnamefont {Tsujikawa}},\
  }\href {\doibase 10.1016/j.physletb.2023.138022} {\bibfield  {journal}
  {\bibinfo  {journal} {\emph {Phys. Lett. B}}\ }\textbf {\bibinfo {volume}
  {843}},\ \bibinfo {pages} {138022} (\bibinfo {year} {2023})},\ \Eprint
  {http://arxiv.org/abs/2303.13717} {arXiv:2303.13717 [gr-qc]} \BibitemShut
  {NoStop}%
\bibitem [{\citenamefont {Bueno}\ and\ \citenamefont
  {Cano}(2016{\natexlab{a}})}]{Bueno:2016xff}%
  \BibitemOpen
  \bibfield  {author} {\bibinfo {author} {\bibfnamefont {P.}~\bibnamefont
  {Bueno}} and \bibinfo {author} {\bibfnamefont {P.~A.}\ \bibnamefont {Cano}},\
  }\href {\doibase 10.1103/PhysRevD.94.104005} {\bibfield  {journal} {\bibinfo
  {journal} {\emph {Phys. Rev. D}}\ }\textbf {\bibinfo {volume} {94}},\
  \bibinfo {pages} {104005} (\bibinfo {year} {2016}{\natexlab{a}})},\ \Eprint
  {http://arxiv.org/abs/1607.06463} {arXiv:1607.06463 [hep-th]} \BibitemShut
  {NoStop}%
\bibitem [{\citenamefont {Bueno}\ \emph {et~al.}(2017)\citenamefont {Bueno},
  \citenamefont {Cano}, \citenamefont {Min},\ and\ \citenamefont
  {Visser}}]{Bueno:2016ypa}%
  \BibitemOpen
  \bibfield  {author} {\bibinfo {author} {\bibfnamefont {P.}~\bibnamefont
  {Bueno}}, \bibinfo {author} {\bibfnamefont {P.~A.}\ \bibnamefont {Cano}},
  \bibinfo {author} {\bibfnamefont {V.~S.}\ \bibnamefont {Min}},  and \bibinfo
  {author} {\bibfnamefont {M.~R.}\ \bibnamefont {Visser}},\ }\href {\doibase
  10.1103/PhysRevD.95.044010} {\bibfield  {journal} {\bibinfo  {journal} {\emph
  {Phys. Rev. D}}\ }\textbf {\bibinfo {volume} {95}},\ \bibinfo {pages}
  {044010} (\bibinfo {year} {2017})},\ \Eprint
  {http://arxiv.org/abs/1610.08519} {arXiv:1610.08519 [hep-th]} \BibitemShut
  {NoStop}%
\bibitem [{\citenamefont {Hennigar}\ and\ \citenamefont
  {Mann}(2017)}]{Hennigar:2016gkm}%
  \BibitemOpen
  \bibfield  {author} {\bibinfo {author} {\bibfnamefont {R.~A.}\ \bibnamefont
  {Hennigar}} and \bibinfo {author} {\bibfnamefont {R.~B.}\ \bibnamefont
  {Mann}},\ }\href {\doibase 10.1103/PhysRevD.95.064055} {\bibfield  {journal}
  {\bibinfo  {journal} {\emph {Phys. Rev. D}}\ }\textbf {\bibinfo {volume}
  {95}},\ \bibinfo {pages} {064055} (\bibinfo {year} {2017})},\ \Eprint
  {http://arxiv.org/abs/1610.06675} {arXiv:1610.06675 [hep-th]} \BibitemShut
  {NoStop}%
\bibitem [{\citenamefont {Bueno}\ and\ \citenamefont
  {Cano}(2016{\natexlab{b}})}]{Bueno:2016lrh}%
  \BibitemOpen
  \bibfield  {author} {\bibinfo {author} {\bibfnamefont {P.}~\bibnamefont
  {Bueno}} and \bibinfo {author} {\bibfnamefont {P.~A.}\ \bibnamefont {Cano}},\
  }\href {\doibase 10.1103/PhysRevD.94.124051} {\bibfield  {journal} {\bibinfo
  {journal} {\emph {Phys. Rev. D}}\ }\textbf {\bibinfo {volume} {94}},\
  \bibinfo {pages} {124051} (\bibinfo {year} {2016}{\natexlab{b}})},\ \Eprint
  {http://arxiv.org/abs/1610.08019} {arXiv:1610.08019 [hep-th]} \BibitemShut
  {NoStop}%
\bibitem [{\citenamefont {Bueno}\ and\ \citenamefont
  {Cano}(2017)}]{Bueno:2017sui}%
  \BibitemOpen
  \bibfield  {author} {\bibinfo {author} {\bibfnamefont {P.}~\bibnamefont
  {Bueno}} and \bibinfo {author} {\bibfnamefont {P.~A.}\ \bibnamefont {Cano}},\
  }\href {\doibase 10.1088/1361-6382/aa8056} {\bibfield  {journal} {\bibinfo
  {journal} {\emph {Class. Quant. Grav.}}\ }\textbf {\bibinfo {volume} {34}},\
  \bibinfo {pages} {175008} (\bibinfo {year} {2017})},\ \Eprint
  {http://arxiv.org/abs/1703.04625} {arXiv:1703.04625 [hep-th]} \BibitemShut
  {NoStop}%
\bibitem [{\citenamefont {Hennigar}\ \emph {et~al.}(2018)\citenamefont
  {Hennigar}, \citenamefont {Poshteh},\ and\ \citenamefont
  {Mann}}]{Hennigar:2018hza}%
  \BibitemOpen
  \bibfield  {author} {\bibinfo {author} {\bibfnamefont {R.~A.}\ \bibnamefont
  {Hennigar}}, \bibinfo {author} {\bibfnamefont {M.~B.~J.}\ \bibnamefont
  {Poshteh}},  and \bibinfo {author} {\bibfnamefont {R.~B.}\ \bibnamefont
  {Mann}},\ }\href {\doibase 10.1103/PhysRevD.97.064041} {\bibfield  {journal}
  {\bibinfo  {journal} {\emph {Phys. Rev. D}}\ }\textbf {\bibinfo {volume}
  {97}},\ \bibinfo {pages} {064041} (\bibinfo {year} {2018})},\ \Eprint
  {http://arxiv.org/abs/1801.03223} {arXiv:1801.03223 [gr-qc]} \BibitemShut
  {NoStop}%
\bibitem [{\citenamefont {Adair}\ \emph {et~al.}(2020)\citenamefont {Adair},
  \citenamefont {Bueno}, \citenamefont {Cano}, \citenamefont {Hennigar},\ and\
  \citenamefont {Mann}}]{Adair:2020vso}%
  \BibitemOpen
  \bibfield  {author} {\bibinfo {author} {\bibfnamefont {C.}~\bibnamefont
  {Adair}}, \bibinfo {author} {\bibfnamefont {P.}~\bibnamefont {Bueno}},
  \bibinfo {author} {\bibfnamefont {P.~A.}\ \bibnamefont {Cano}}, \bibinfo
  {author} {\bibfnamefont {R.~A.}\ \bibnamefont {Hennigar}},  and \bibinfo
  {author} {\bibfnamefont {R.~B.}\ \bibnamefont {Mann}},\ }\href {\doibase
  10.1103/PhysRevD.102.084001} {\bibfield  {journal} {\bibinfo  {journal}
  {\emph {Phys. Rev. D}}\ }\textbf {\bibinfo {volume} {102}},\ \bibinfo {pages}
  {084001} (\bibinfo {year} {2020})},\ \Eprint
  {http://arxiv.org/abs/2004.09598} {arXiv:2004.09598 [gr-qc]} \BibitemShut
  {NoStop}%
\bibitem [{\citenamefont {Frassino}\ and\ \citenamefont
  {Rocha}(2020)}]{Frassino:2020zuv}%
  \BibitemOpen
  \bibfield  {author} {\bibinfo {author} {\bibfnamefont {A.~M.}\ \bibnamefont
  {Frassino}} and \bibinfo {author} {\bibfnamefont {J.~V.}\ \bibnamefont
  {Rocha}},\ }\href {\doibase 10.1103/PhysRevD.102.024035} {\bibfield
  {journal} {\bibinfo  {journal} {\emph {Phys. Rev. D}}\ }\textbf {\bibinfo
  {volume} {102}},\ \bibinfo {pages} {024035} (\bibinfo {year} {2020})},\
  \Eprint {http://arxiv.org/abs/2002.04071} {arXiv:2002.04071 [hep-th]}
  \BibitemShut {NoStop}%
\bibitem [{\citenamefont {Burger}\ \emph {et~al.}(2020)\citenamefont {Burger},
  \citenamefont {Emond},\ and\ \citenamefont {Moynihan}}]{Burger:2019wkq}%
  \BibitemOpen
  \bibfield  {author} {\bibinfo {author} {\bibfnamefont {D.~J.}\ \bibnamefont
  {Burger}}, \bibinfo {author} {\bibfnamefont {W.~T.}\ \bibnamefont {Emond}},
  and \bibinfo {author} {\bibfnamefont {N.}~\bibnamefont {Moynihan}},\ }\href
  {\doibase 10.1103/PhysRevD.101.084009} {\bibfield  {journal} {\bibinfo
  {journal} {\emph {Phys. Rev. D}}\ }\textbf {\bibinfo {volume} {101}},\
  \bibinfo {pages} {084009} (\bibinfo {year} {2020})},\ \Eprint
  {http://arxiv.org/abs/1910.11618} {arXiv:1910.11618 [hep-th]} \BibitemShut
  {NoStop}%
\bibitem [{\citenamefont {Cano}\ and\ \citenamefont
  {Pere\~niguez}(2020)}]{Cano:2019ozf}%
  \BibitemOpen
  \bibfield  {author} {\bibinfo {author} {\bibfnamefont {P.~A.}\ \bibnamefont
  {Cano}} and \bibinfo {author} {\bibfnamefont {D.}~\bibnamefont
  {Pere\~niguez}},\ }\href {\doibase 10.1103/PhysRevD.101.044016} {\bibfield
  {journal} {\bibinfo  {journal} {\emph {Phys. Rev. D}}\ }\textbf {\bibinfo
  {volume} {101}},\ \bibinfo {pages} {044016} (\bibinfo {year} {2020})},\
  \Eprint {http://arxiv.org/abs/1910.10721} {arXiv:1910.10721 [hep-th]}
  \BibitemShut {NoStop}%
\bibitem [{\citenamefont {Hennigar}\ \emph {et~al.}(2017)\citenamefont
  {Hennigar}, \citenamefont {Kubiz\v{n}\'ak},\ and\ \citenamefont
  {Mann}}]{Hennigar:2017ego}%
  \BibitemOpen
  \bibfield  {author} {\bibinfo {author} {\bibfnamefont {R.~A.}\ \bibnamefont
  {Hennigar}}, \bibinfo {author} {\bibfnamefont {D.}~\bibnamefont
  {Kubiz\v{n}\'ak}},  and \bibinfo {author} {\bibfnamefont {R.~B.}\
  \bibnamefont {Mann}},\ }\href {\doibase 10.1103/PhysRevD.95.104042}
  {\bibfield  {journal} {\bibinfo  {journal} {\emph {Phys. Rev. D}}\ }\textbf
  {\bibinfo {volume} {95}},\ \bibinfo {pages} {104042} (\bibinfo {year}
  {2017})},\ \Eprint {http://arxiv.org/abs/1703.01631} {arXiv:1703.01631
  [hep-th]} \BibitemShut {NoStop}%
\bibitem [{\citenamefont {Bueno}\ \emph {et~al.}(2019)\citenamefont {Bueno},
  \citenamefont {Cano}, \citenamefont {Moreno},\ and\ \citenamefont
  {Murcia}}]{Bueno:2019ltp}%
  \BibitemOpen
  \bibfield  {author} {\bibinfo {author} {\bibfnamefont {P.}~\bibnamefont
  {Bueno}}, \bibinfo {author} {\bibfnamefont {P.~A.}\ \bibnamefont {Cano}},
  \bibinfo {author} {\bibfnamefont {J.}~\bibnamefont {Moreno}},  and \bibinfo
  {author} {\bibfnamefont {A.}~\bibnamefont {Murcia}},\ }\href {\doibase
  10.1007/JHEP11(2019)062} {\bibfield  {journal} {\bibinfo  {journal} {\emph
  {JHEP}}\ }\textbf {\bibinfo {volume} {11}},\ \bibinfo {pages} {062} (\bibinfo
  {year} {2019})},\ \Eprint {http://arxiv.org/abs/1906.00987} {arXiv:1906.00987
  [hep-th]} \BibitemShut {NoStop}%
\bibitem [{\citenamefont {Bueno}\ \emph {et~al.}(2020)\citenamefont {Bueno},
  \citenamefont {Cano},\ and\ \citenamefont {Hennigar}}]{Bueno:2019ycr}%
  \BibitemOpen
  \bibfield  {author} {\bibinfo {author} {\bibfnamefont {P.}~\bibnamefont
  {Bueno}}, \bibinfo {author} {\bibfnamefont {P.~A.}\ \bibnamefont {Cano}},
  and \bibinfo {author} {\bibfnamefont {R.~A.}\ \bibnamefont {Hennigar}},\
  }\href {\doibase 10.1088/1361-6382/ab5410} {\bibfield  {journal} {\bibinfo
  {journal} {\emph {Class. Quant. Grav.}}\ }\textbf {\bibinfo {volume} {37}},\
  \bibinfo {pages} {015002} (\bibinfo {year} {2020})},\ \Eprint
  {http://arxiv.org/abs/1909.07983} {arXiv:1909.07983 [hep-th]} \BibitemShut
  {NoStop}%
\bibitem [{\citenamefont {Arciniega}\ \emph
  {et~al.}(2020{\natexlab{a}})\citenamefont {Arciniega}, \citenamefont {Bueno},
  \citenamefont {Cano}, \citenamefont {Edelstein}, \citenamefont {Hennigar},\
  and\ \citenamefont {Jaime}}]{Arciniega:2018tnn}%
  \BibitemOpen
  \bibfield  {author} {\bibinfo {author} {\bibfnamefont {G.}~\bibnamefont
  {Arciniega}}, \bibinfo {author} {\bibfnamefont {P.}~\bibnamefont {Bueno}},
  \bibinfo {author} {\bibfnamefont {P.~A.}\ \bibnamefont {Cano}}, \bibinfo
  {author} {\bibfnamefont {J.~D.}\ \bibnamefont {Edelstein}}, \bibinfo {author}
  {\bibfnamefont {R.~A.}\ \bibnamefont {Hennigar}},  and \bibinfo {author}
  {\bibfnamefont {L.~G.}\ \bibnamefont {Jaime}},\ }\href {\doibase
  10.1016/j.physletb.2020.135242} {\bibfield  {journal} {\bibinfo  {journal}
  {\emph {Phys. Lett. B}}\ }\textbf {\bibinfo {volume} {802}},\ \bibinfo
  {pages} {135242} (\bibinfo {year} {2020}{\natexlab{a}})},\ \Eprint
  {http://arxiv.org/abs/1812.11187} {arXiv:1812.11187 [hep-th]} \BibitemShut
  {NoStop}%
\bibitem [{\citenamefont {Arciniega}\ \emph
  {et~al.}(2020{\natexlab{b}})\citenamefont {Arciniega}, \citenamefont
  {Edelstein},\ and\ \citenamefont {Jaime}}]{Arciniega:2018fxj}%
  \BibitemOpen
  \bibfield  {author} {\bibinfo {author} {\bibfnamefont {G.}~\bibnamefont
  {Arciniega}}, \bibinfo {author} {\bibfnamefont {J.~D.}\ \bibnamefont
  {Edelstein}},  and \bibinfo {author} {\bibfnamefont {L.~G.}\ \bibnamefont
  {Jaime}},\ }\href {\doibase 10.1016/j.physletb.2020.135272} {\bibfield
  {journal} {\bibinfo  {journal} {\emph {Phys. Lett. B}}\ }\textbf {\bibinfo
  {volume} {802}},\ \bibinfo {pages} {135272} (\bibinfo {year}
  {2020}{\natexlab{b}})},\ \Eprint {http://arxiv.org/abs/1810.08166}
  {arXiv:1810.08166 [gr-qc]} \BibitemShut {NoStop}%
\bibitem [{\citenamefont {Cisterna}\ \emph {et~al.}(2020)\citenamefont
  {Cisterna}, \citenamefont {Grandi},\ and\ \citenamefont
  {Oliva}}]{Cisterna:2018tgx}%
  \BibitemOpen
  \bibfield  {author} {\bibinfo {author} {\bibfnamefont {A.}~\bibnamefont
  {Cisterna}}, \bibinfo {author} {\bibfnamefont {N.}~\bibnamefont {Grandi}},
  and \bibinfo {author} {\bibfnamefont {J.}~\bibnamefont {Oliva}},\ }\href
  {\doibase 10.1016/j.physletb.2020.135435} {\bibfield  {journal} {\bibinfo
  {journal} {\emph {Phys. Lett. B}}\ }\textbf {\bibinfo {volume} {805}},\
  \bibinfo {pages} {135435} (\bibinfo {year} {2020})},\ \Eprint
  {http://arxiv.org/abs/1811.06523} {arXiv:1811.06523 [hep-th]} \BibitemShut
  {NoStop}%
\bibitem [{\citenamefont {Erices}\ \emph {et~al.}(2019)\citenamefont {Erices},
  \citenamefont {Papantonopoulos},\ and\ \citenamefont
  {Saridakis}}]{Erices:2019mkd}%
  \BibitemOpen
  \bibfield  {author} {\bibinfo {author} {\bibfnamefont {C.}~\bibnamefont
  {Erices}}, \bibinfo {author} {\bibfnamefont {E.}~\bibnamefont
  {Papantonopoulos}},  and \bibinfo {author} {\bibfnamefont {E.~N.}\
  \bibnamefont {Saridakis}},\ }\href {\doibase 10.1103/PhysRevD.99.123527}
  {\bibfield  {journal} {\bibinfo  {journal} {\emph {Phys. Rev. D}}\ }\textbf
  {\bibinfo {volume} {99}},\ \bibinfo {pages} {123527} (\bibinfo {year}
  {2019})},\ \Eprint {http://arxiv.org/abs/1903.11128} {arXiv:1903.11128
  [gr-qc]} \BibitemShut {NoStop}%
\bibitem [{\citenamefont {Quiros}\ \emph {et~al.}(2020)\citenamefont {Quiros},
  \citenamefont {Garc\'\i{}a-Salcedo}, \citenamefont {Gonzalez}, \citenamefont
  {Mart\'\i{}nez},\ and\ \citenamefont {Nucamendi}}]{Quiros:2020uhr}%
  \BibitemOpen
  \bibfield  {author} {\bibinfo {author} {\bibfnamefont {I.}~\bibnamefont
  {Quiros}}, \bibinfo {author} {\bibfnamefont {R.}~\bibnamefont
  {Garc\'\i{}a-Salcedo}}, \bibinfo {author} {\bibfnamefont {T.}~\bibnamefont
  {Gonzalez}}, \bibinfo {author} {\bibfnamefont {J.~L.~M.}\ \bibnamefont
  {Mart\'\i{}nez}},  and \bibinfo {author} {\bibfnamefont {U.}~\bibnamefont
  {Nucamendi}},\ }\href {\doibase 10.1103/PhysRevD.102.044018} {\bibfield
  {journal} {\bibinfo  {journal} {\emph {Phys. Rev. D}}\ }\textbf {\bibinfo
  {volume} {102}},\ \bibinfo {pages} {044018} (\bibinfo {year} {2020})},\
  \Eprint {http://arxiv.org/abs/2003.10516} {arXiv:2003.10516 [gr-qc]}
  \BibitemShut {NoStop}%
\bibitem [{\citenamefont {Beltr\'an~Jim\'enez}\ and\ \citenamefont
  {Jim\'enez-Cano}(2021)}]{BeltranJimenez:2020lee}%
  \BibitemOpen
  \bibfield  {author} {\bibinfo {author} {\bibfnamefont {J.}~\bibnamefont
  {Beltr\'an~Jim\'enez}} and \bibinfo {author} {\bibfnamefont {A.}~\bibnamefont
  {Jim\'enez-Cano}},\ }\href {\doibase 10.1088/1475-7516/2021/01/069}
  {\bibfield  {journal} {\bibinfo  {journal} {\emph {JCAP}}\ }\textbf {\bibinfo
  {volume} {01}},\ \bibinfo {pages} {069} (\bibinfo {year} {2021})},\ \Eprint
  {http://arxiv.org/abs/2009.08197} {arXiv:2009.08197 [gr-qc]} \BibitemShut
  {NoStop}%
\bibitem [{\citenamefont {Cano}\ \emph {et~al.}(2021)\citenamefont {Cano},
  \citenamefont {Fransen},\ and\ \citenamefont {Hertog}}]{Cano:2020oaa}%
  \BibitemOpen
  \bibfield  {author} {\bibinfo {author} {\bibfnamefont {P.~A.}\ \bibnamefont
  {Cano}}, \bibinfo {author} {\bibfnamefont {K.}~\bibnamefont {Fransen}},  and
  \bibinfo {author} {\bibfnamefont {T.}~\bibnamefont {Hertog}},\ }\href
  {\doibase 10.1103/PhysRevD.103.103531} {\bibfield  {journal} {\bibinfo
  {journal} {\emph {Phys. Rev. D}}\ }\textbf {\bibinfo {volume} {103}},\
  \bibinfo {pages} {103531} (\bibinfo {year} {2021})},\ \Eprint
  {http://arxiv.org/abs/2011.13933} {arXiv:2011.13933 [hep-th]} \BibitemShut
  {NoStop}%
\bibitem [{\citenamefont {Pookkillath}\ \emph {et~al.}(2020)\citenamefont
  {Pookkillath}, \citenamefont {De~Felice},\ and\ \citenamefont
  {Starobinsky}}]{Pookkillath:2020iqq}%
  \BibitemOpen
  \bibfield  {author} {\bibinfo {author} {\bibfnamefont {M.~C.}\ \bibnamefont
  {Pookkillath}}, \bibinfo {author} {\bibfnamefont {A.}~\bibnamefont
  {De~Felice}},  and \bibinfo {author} {\bibfnamefont {A.~A.}\ \bibnamefont
  {Starobinsky}},\ }\href {\doibase 10.1088/1475-7516/2020/07/041} {\bibfield
  {journal} {\bibinfo  {journal} {\emph {JCAP}}\ }\textbf {\bibinfo {volume}
  {07}},\ \bibinfo {pages} {041} (\bibinfo {year} {2020})},\ \Eprint
  {http://arxiv.org/abs/2004.03912} {arXiv:2004.03912 [gr-qc]} \BibitemShut
  {NoStop}%
\bibitem [{\citenamefont {Regge}\ and\ \citenamefont
  {Wheeler}(1957)}]{Regge:1957td}%
  \BibitemOpen
  \bibfield  {author} {\bibinfo {author} {\bibfnamefont {T.}~\bibnamefont
  {Regge}} and \bibinfo {author} {\bibfnamefont {J.~A.}\ \bibnamefont
  {Wheeler}},\ }\href {\doibase 10.1103/PhysRev.108.1063} {\bibfield  {journal}
  {\bibinfo  {journal} {\emph {Phys. Rev.}}\ }\textbf {\bibinfo {volume}
  {108}},\ \bibinfo {pages} {1063} (\bibinfo {year} {1957})}\BibitemShut
  {NoStop}%
\bibitem [{\citenamefont {Jim\'enez}\ and\ \citenamefont
  {Jim\'enez-Cano}(2023)}]{Jimenez:2023esa}%
  \BibitemOpen
  \bibfield  {author} {\bibinfo {author} {\bibfnamefont {J.~B.}\ \bibnamefont
  {Jim\'enez}} and \bibinfo {author} {\bibfnamefont {A.}~\bibnamefont
  {Jim\'enez-Cano}},\ }\Eprint {http://arxiv.org/abs/2306.07095}
  {arXiv:2306.07095 [gr-qc]} \BibitemShut {NoStop}%
\bibitem [{\citenamefont {Starobinsky}(1980)}]{Starobinsky:1980te}%
  \BibitemOpen
  \bibfield  {author} {\bibinfo {author} {\bibfnamefont {A.~A.}\ \bibnamefont
  {Starobinsky}},\ }\href {\doibase 10.1016/0370-2693(80)90670-X} {\bibfield
  {journal} {\bibinfo  {journal} {\emph {Phys. Lett. B}}\ }\textbf {\bibinfo
  {volume} {91}},\ \bibinfo {pages} {99} (\bibinfo {year} {1980})}\BibitemShut
  {NoStop}%
\bibitem [{\citenamefont {Zerilli}(1970)}]{Zerilli:1970se}%
  \BibitemOpen
  \bibfield  {author} {\bibinfo {author} {\bibfnamefont {F.~J.}\ \bibnamefont
  {Zerilli}},\ }\href {\doibase 10.1103/PhysRevLett.24.737} {\bibfield
  {journal} {\bibinfo  {journal} {\emph {Phys. Rev. Lett.}}\ }\textbf {\bibinfo
  {volume} {24}},\ \bibinfo {pages} {737} (\bibinfo {year} {1970})}\BibitemShut
  {NoStop}%
\bibitem [{\citenamefont {De~Felice}\ \emph {et~al.}(2011)\citenamefont
  {De~Felice}, \citenamefont {Suyama},\ and\ \citenamefont
  {Tanaka}}]{DeFelice:2011ka}%
  \BibitemOpen
  \bibfield  {author} {\bibinfo {author} {\bibfnamefont {A.}~\bibnamefont
  {De~Felice}}, \bibinfo {author} {\bibfnamefont {T.}~\bibnamefont {Suyama}},
  and \bibinfo {author} {\bibfnamefont {T.}~\bibnamefont {Tanaka}},\ }\href
  {\doibase 10.1103/PhysRevD.83.104035} {\bibfield  {journal} {\bibinfo
  {journal} {\emph {Phys. Rev. D}}\ }\textbf {\bibinfo {volume} {83}},\
  \bibinfo {pages} {104035} (\bibinfo {year} {2011})},\ \Eprint
  {http://arxiv.org/abs/1102.1521} {arXiv:1102.1521 [gr-qc]} \BibitemShut
  {NoStop}%
\bibitem [{\citenamefont {Motohashi}\ and\ \citenamefont
  {Suyama}(2011)}]{Motohashi:2011pw}%
  \BibitemOpen
  \bibfield  {author} {\bibinfo {author} {\bibfnamefont {H.}~\bibnamefont
  {Motohashi}} and \bibinfo {author} {\bibfnamefont {T.}~\bibnamefont
  {Suyama}},\ }\href {\doibase 10.1103/PhysRevD.84.084041} {\bibfield
  {journal} {\bibinfo  {journal} {\emph {Phys. Rev. D}}\ }\textbf {\bibinfo
  {volume} {84}},\ \bibinfo {pages} {084041} (\bibinfo {year} {2011})},\
  \Eprint {http://arxiv.org/abs/1107.3705} {arXiv:1107.3705 [gr-qc]}
  \BibitemShut {NoStop}%
\bibitem [{\citenamefont {Kobayashi}\ \emph {et~al.}(2012)\citenamefont
  {Kobayashi}, \citenamefont {Motohashi},\ and\ \citenamefont
  {Suyama}}]{Kobayashi:2012kh}%
  \BibitemOpen
  \bibfield  {author} {\bibinfo {author} {\bibfnamefont {T.}~\bibnamefont
  {Kobayashi}}, \bibinfo {author} {\bibfnamefont {H.}~\bibnamefont
  {Motohashi}},  and \bibinfo {author} {\bibfnamefont {T.}~\bibnamefont
  {Suyama}},\ }\href {\doibase 10.1103/PhysRevD.85.084025} {\bibfield
  {journal} {\bibinfo  {journal} {\emph {Phys. Rev. D}}\ }\textbf {\bibinfo
  {volume} {85}},\ \bibinfo {pages} {084025} (\bibinfo {year} {2012})},\
  \bibinfo {note} {[Erratum:
  \href{https://doi.org/10.1103/PhysRevD.96.109903}{{\it Phys. Rev. D} {\bf
  96}, 109903 (2017)}]},\ \Eprint {http://arxiv.org/abs/1202.4893}
  {arXiv:1202.4893 [gr-qc]} \BibitemShut {NoStop}%
\bibitem [{\citenamefont {Kase}\ \emph {et~al.}(2014)\citenamefont {Kase},
  \citenamefont {Gergely},\ and\ \citenamefont {Tsujikawa}}]{Kase:2014baa}%
  \BibitemOpen
  \bibfield  {author} {\bibinfo {author} {\bibfnamefont {R.}~\bibnamefont
  {Kase}}, \bibinfo {author} {\bibfnamefont {L.~A.}\ \bibnamefont {Gergely}},
  and \bibinfo {author} {\bibfnamefont {S.}~\bibnamefont {Tsujikawa}},\ }\href
  {\doibase 10.1103/PhysRevD.90.124019} {\bibfield  {journal} {\bibinfo
  {journal} {\emph {Phys. Rev. D}}\ }\textbf {\bibinfo {volume} {90}},\
  \bibinfo {pages} {124019} (\bibinfo {year} {2014})},\ \Eprint
  {http://arxiv.org/abs/1406.2402} {arXiv:1406.2402 [hep-th]} \BibitemShut
  {NoStop}%
\bibitem [{\citenamefont {Kase}\ and\ \citenamefont
  {Tsujikawa}(2022)}]{Kase:2021mix}%
  \BibitemOpen
  \bibfield  {author} {\bibinfo {author} {\bibfnamefont {R.}~\bibnamefont
  {Kase}} and \bibinfo {author} {\bibfnamefont {S.}~\bibnamefont {Tsujikawa}},\
  }\href {\doibase 10.1103/PhysRevD.105.024059} {\bibfield  {journal} {\bibinfo
   {journal} {\emph {Phys. Rev. D}}\ }\textbf {\bibinfo {volume} {105}},\
  \bibinfo {pages} {024059} (\bibinfo {year} {2022})},\ \Eprint
  {http://arxiv.org/abs/2110.12728} {arXiv:2110.12728 [gr-qc]} \BibitemShut
  {NoStop}%
\bibitem [{\citenamefont {Bueno}\ \emph {et~al.}(2023)\citenamefont {Bueno},
  \citenamefont {Cano},\ and\ \citenamefont {Hennigar}}]{Bueno:2023jtc}%
  \BibitemOpen
  \bibfield  {author} {\bibinfo {author} {\bibfnamefont {P.}~\bibnamefont
  {Bueno}}, \bibinfo {author} {\bibfnamefont {P.~A.}\ \bibnamefont {Cano}},
  and \bibinfo {author} {\bibfnamefont {R.~A.}\ \bibnamefont {Hennigar}},\
  }\Eprint {http://arxiv.org/abs/2306.02924} {arXiv:2306.02924 [hep-th]}
  \BibitemShut {NoStop}%
\end{thebibliography}%

\end{document}